\documentclass[conference]{IEEEtran}
\IEEEoverridecommandlockouts

\usepackage[numbers,sort&compress]{natbib}
\usepackage{amsmath,amssymb,amsfonts}
\usepackage{algorithmic}
\usepackage{graphicx}
\usepackage{textcomp}
\usepackage[table,xcdraw]{xcolor}
\usepackage[hyphens]{url}
\usepackage[linesnumbered,ruled,vlined]{algorithm2e}
\usepackage[inline]{enumitem}
\usepackage{algorithmic}
\usepackage{pgfplots}
\usepackage{todonotes}
\usepackage{svg}
\usepackage{subcaption}
\usepackage{multicol}
\usepackage{multirow}
\usepackage{adjustbox}

\def\BibTeX{{\rm B\kern-.05em{\sc i\kern-.025em b}\kern-.08em
    T\kern-.1667em\lower.7ex\hbox{E}\kern-.125emX}}
\begin{document}

\pdfpagewidth=8.5in
\pdfpageheight=11in

\pagenumbering{arabic}

\title{ORAP: \textbf{O}ptimized \textbf{R}ow \textbf{A}ccess \textbf{P}refetching for Rowhammer-mitigated Memory}

\author{\IEEEauthorblockN{Maccoy Merrell\IEEEauthorrefmark{1},
Daniel Puckett\IEEEauthorrefmark{1}, Gino Chacon\IEEEauthorrefmark{2},
Jeffrey Stuecheli\IEEEauthorrefmark{3}, Stavros Kalafatis\IEEEauthorrefmark{1}, Paul V. Gratz\IEEEauthorrefmark{1}}

\begin{tabular}{c c c}
\IEEEauthorblockA{
\IEEEauthorrefmark{1}Department of Electrical and Computer Engineering & \IEEEauthorrefmark{2}AheadComputing & \IEEEauthorrefmark{3}Arm\\
Texas A\&M University, College Station, TX, USA & Beaverton, OR, USA & Austin, TX, USA
}
\end{tabular}
\IEEEauthorblockA{
\{maccoy.merrell, dpuckett98, skalafatis-tamu, pgratz\}@tamu.edu \\ ginoachacon@gmail.com jeff.stuecheli@arm.com}
}

\maketitle
\thispagestyle{plain}
\pagestyle{plain}

\begin{abstract}

Rowhammer is a well-studied DRAM phenomenon wherein
  multiple activations to a given row can cause bit flips in adjacent
  rows.  Many mitigation techniques have also been introduced to
  address Rowhammer, with some support being already incorporated into
  the JEDEC DDR5 standard, specifically for per-row-activation-counter
  (PRAC) and refresh-management (RFM) systems. Mitigation schemes built
  on these mechanisms claim to have various levels of area, power, and
  performance overheads.  To date the evaluation of existing
  mitigation schemes typically neglects the impact of other memory
  system components such as hardware prefetchers.  Nearly all modern
  systems incorporate multiple levels of hardware prefetching and
  these can significantly improve processor performance
  through speculative cache population.  While hiding memory
  latencies, these prefetchers induce higher numbers of downstream
  memory requests and increase DRAM activation rates. The performance
  overhead of Rowhammer mitigations are tied directly to memory access
  patterns, exposing both hardware prefetchers and Rowhammer
  mitigations to cross-interaction.  As a result, we find that the
  performance improvement provided by prior-work hardware prefetchers
  is often severely impacted by typical Rowhammer mitigations.  In
  effect, much of the benefit of speculative memory references from
  prefetching lies in accelerating and reordering DRAM references in
  ways that can trigger mitigations, incurring extra latency and
  significantly reducing the benefits of prefetching.

  This work proposes the Optimized Row Access Prefetcher (ORAP),
  leveraging abundant last-level-cache (LLC) space to cache large
  portions of DRAM rowbuffer contents to reduce the need for future
  activations. Working alongside the state-of-the-art Berti prefetcher 
  in the first level caches, ORAP reduces DRAM activation rates by 51.3\% 
  and achieves a 4.6\% speedup over the state-of-the-art prefetcher
  configuration of Berti and SPP-PPF when prefetching in an RFM-mitigated memory system.
  Under PRAC mitigations, ORAP reduces energy overheads by 11.8\%.

\end{abstract}

\section{Introduction}
Rowhammer is a DRAM hardware vulnerability first discovered by Kim et
al. \cite{flipping-bits-in-memory} and further characterized in
subsequent works \cite{revisiting-rowhammer,
  rowhammer-a-retrospective, the-rowhammer-problem}.  Early works
focused on how this could be leveraged to carry out various attacks
\cite{rambleed, memway, sledgehammer}, while contemporary research
focuses primarily on developing or bypassing hardware
mitigations. Many mitigation techniques have been proposed
\cite{mithril,hydra,mint,blockhammer,randomized-rowswap,graphene},
with strategies like targeted-row-refresh (TRR) having been adopted
for use in industry \cite{trespass,zen4map}. The JEDEC DDR5 standard
currently supports some of these mitigations by standardizing
refresh-management (RFM)~\cite{rfm} commands and the use of
per-row-activation-counters (PRAC)~\cite{prac1,prac2}. While these
mitigation techniques broadly are able to reduce the likelihood of
Rowhammer bit flips, this can come at some cost to performance and/or
other overheads.  Mitigation performance costs come in the form of
higher DRAM latency and energy, particularly for memory-bound workloads.

Hardware prefetching is a well-studied and heavily utilized technique
to improve performance \cite{ppf,berti,bingo,ampm,bop,ipcp}, where a
prefetch engine within one of the CPU's data caches learns and
speculatively fetches cache lines that it believes will be used in the
future.  State-of-the-art prefetching systems display a 21.5\%
multi-core speedup over non-prefetching systems, a significant
performance improvement for relatively little state.

\begin{figure}[]
	\centering
	\includegraphics[width=1.0\columnwidth]{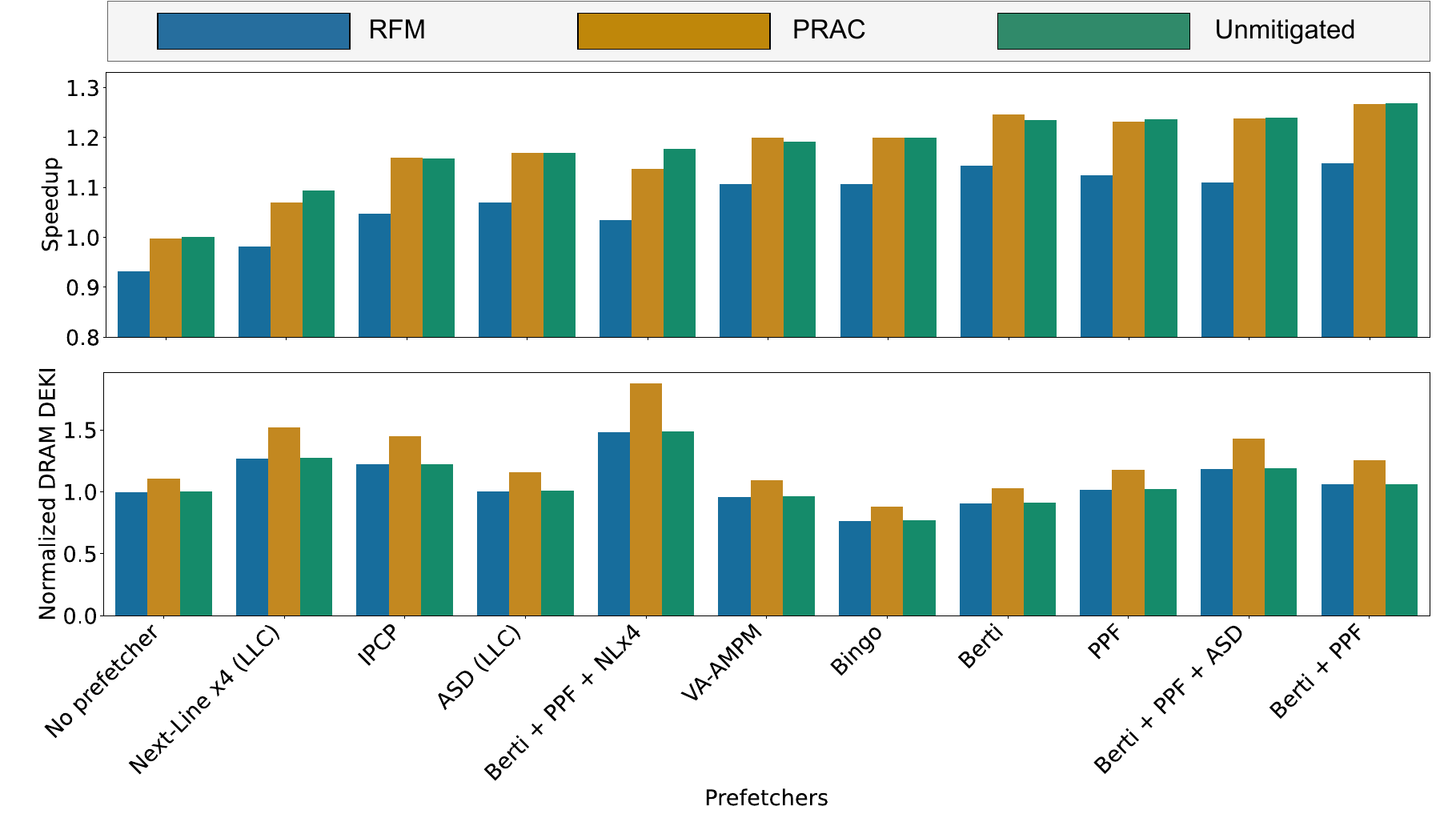}
        \caption{Impact of Rowhammer mitigations on prefetcher single-core performance improvement (top) and DRAM dynamic energy per thousand instructions (bottom)}
        \label{fig:mitigation_effect_on_prefetchers}
\end{figure}

Despite both Rowhammer mitigations and hardware prefetching being
typically implemented in most commercial systems, generally the
interactions between them have not been studied.  Hardware prefetchers
excel at reducing apparent memory latency, but are also known to
induce additional memory requests over non-prefetching systems. The
increased stress on the memory system is considered acceptable in
academic works as the performance improvement is far more
significant. Mitigation overheads, evaluated without these
prefetchers, see an unrealistic system whose existing memory streams
are well-ordered and efficient with heavy memory bottlenecks. With
prefetchers already included in nearly all shipping systems and
Rowhammer mitigations seeing widespread adoption as systems are
upgraded to DDR5 memories, these interactions are actively impacting
real-world devices and are yet-to-be
studied. Figure~\ref{fig:mitigation_effect_on_prefetchers} shows the
single-core geomean speedup of different prefetchers on Rowhammer
mitigated systems alongside DRAM dynamic energy. We find that the 
speedup and energy costs of these policies are reliant on 
the mitigation used. In particular, RFM significantly reduces the performance benefit of prefetchers, while PRAC significantly increases their energy overhead.

In this work, we study the interactions of hardware prefetchers and
Rowhammer mitigations.  As described above, we find that in Rowhammer
mitigated systems, prefetchers naturally induce more activations and
hence trigger the mitigations more frequently, in turn inducing
significant slowdowns and energy costs. To address this issue we propose 
the Optimized Row Access Prefetcher (ORAP), a last-level cache hardware 
prefetcher designed to reduce the negative interactions between the two systems,
improving performance while lowering row activation counts. This work
makes the following contributions:
\begin{enumerate}
\item We propose ORAP, a Rowhammer-mitigation- and DRAM-aware
  spatial data prefetcher within the LLC, designed to be used
  alongside other prefetchers in the L1 and L2
  caches.
\item Unlike previous works in DRAM-aware prefetching, ORAP
  proactively avoids triggering DRAM activations through deep forward
  speculation and cache management. This is achieved without the
  addition of any new hardware in the memory controller or within the
  DRAM.
\item We develop a novel feedback technique for prefetching into large
  caches, prefetching with delayed usefulness
  information, and preventing excessive cache pollution.
\item ORAP increases the performance of RFM-based (that is, low-cost)
  Rowhammer-mitigated systems over the state-of-the-art by 4.6\%
  without sacrificing security of the underlying
  RFM scheme and 3.3\% for unmitigated systems. Under PRAC-based (high-cost)
  mitigations, DRAM dynamic energy is reduced by 11.8\%.
\end{enumerate}

\section{Background and Motivation} \label{sec:background}
\subsection{DRAM Organization}
\begin{figure}[]
	\centering
	\includesvg[width=1.0\columnwidth]{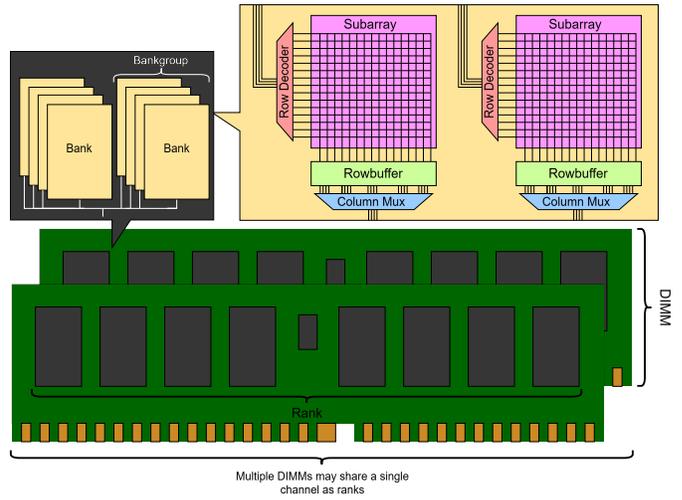}
        \caption{The hierarchical organization of a dual in-line memory module (DIMM)}
        \label{fig:dram_layout}
\end{figure}

DRAM serves as main memory in nearly all modern computing. 
Capacitors paired with guard transistors store
individual bits in large arrays. These capacitors lose
charge over time, so rewrites (refreshes) are required to
preserve memory integrity. Since system memory requirements vary, DRAM
is typically populated within systems as needed, leading to wide variations
in addressable space and available memory bandwidth. While
appearing as a contiguous piece of memory to programmers,
memory systems have subdivided DRAM into multiple levels to support
these many different potential configurations. Figure~\ref{fig:dram_layout} 
shows how these levels correlate to the physical devices. From within 
processors, memory controllers are duplicated per channel, each channel 
being operated independently and utilizing DDR protocols to transmit data on both
clock edges. Within each channel, ranks of DRAM chips which form full
words are addressed individually. Each of these ranks hosts dram banks
which can operate in parallel, allowing multiple requests to be served
simultaneously, though having to reserve a single channel shared with
other ranks for the transmission of data. Each bank is addressed by a
row and column address, which identifies which internal subarray is
active and which bits of that subarray are read or written.

Reads and writes are issued at the bank level from/to the DRAM arrays,
These operations are destructive and slow, so each bank instead acts upon entire rows at once, reading them into internal buffers called rowbuffers which can coalesce multiple DRAM operations before being written back to the array. While more complex than described here, activation commands (ACTs) to the DRAM
serve as reads from the DRAM arrays into these rowbuffers, while
precharge commands (PREs) serve as writes from these rowbuffers back
into the arrays. Rowbuffers within a bank can only store a single row at a time, so it is far more performant to reaccess a rowbuffer with the target row than initiate a data transfer between the rowbuffer and array. Some mitigation schemes such as PRAC increase the time it takes to carry out PRE commands, which in turn can worsen energy consumption.

\subsection{DRAM Address Mappings and Bank-level Parallelism}
\begin{figure}[]
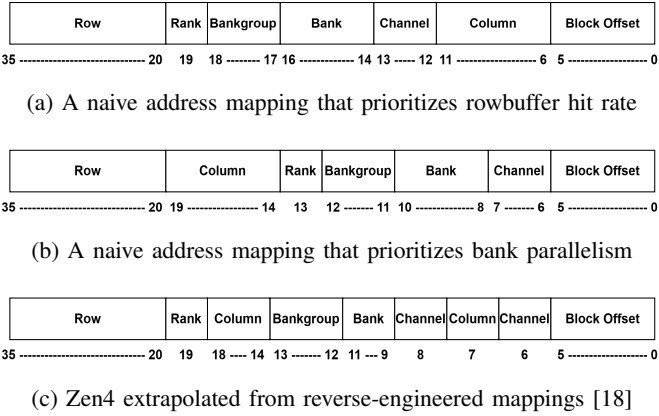

    \centering
    \begin{subfigure}[h!]{1.0\columnwidth}
    \centering
	\includesvg[width=1.0\columnwidth]{figures/hitrate_mapping.svg}\caption{A naive address mapping that prioritizes rowbuffer hit rate}\label{fig:hitrate_mapping}
    \end{subfigure}
    \par\bigskip
    \begin{subfigure}[]{1.0\columnwidth}
    \centering
	\includesvg[width=1.0\columnwidth]{figures/blp_mapping.svg}\caption{A naive address mapping that prioritizes bank parallelism}\label{fig:blp_mapping}
    \end{subfigure}
    \par\bigskip
    \begin{subfigure}[]{1.0\columnwidth}
    \centering
	\includesvg[width=1.0\columnwidth]{figures/zen4_mapping.svg}\caption{Zen4 extrapolated from reverse-engineered mappings~\cite{zen4map}}\label{fig:zen4_mapping}
    \end{subfigure}
    \par\bigskip
    \caption{64 GiB DRAM logical-physical mappings}
    \label{fig:mappings}
\end{figure}

Balancing the cost of rowbuffer misses while
maintaining maximum bank-level-parallelism (BLP) is a tradeoff managed
by the memory controller and its address mappings. Logical-physical
address mappings are designed by memory controller architects to offer
the best performance to many different configurations by maximizing both BLP
and rowbuffer hit rate simultaneously. To
increase the chance of rowbuffer hits, some column addressing bits are
kept low within the address space (referred to here-on as \textit{column cluster bits}) 
while row addressing bits are kept high. The remaining placement 
of the bits for channels, banks, bankgroups, and ranks 
rely on how memory architects have optimized their
system. Placing these bits as low as possible like in
Figure~\ref{fig:blp_mapping} maximizes the available BLP at the cost
of bank access latency and rowbuffer hit rate, while
placing these bits higher like in Figure~\ref{fig:hitrate_mapping}
sacrifices BLP for instantaneous bank access latency and a higher
rowbuffer hit rate.

The placement of these bank indexing bits varies between different
processor manufacturers and technology generations, though they follow
the general strategy of bisecting the column bits with the bank
indexing~\cite{minpage}, exploiting spatial locality for increased
rowbuffer hits without disrupting available BLP. To further maximize
BLP and avoid collisions, row bits are typically XORed with these bank
index bits, resulting in different rows having different bank-wise
strides. Recently, AMD's Zen4 address mappings for their DDR5 memory
controllers were reverse-engineered, showing that these techniques are
still in use~\cite{zen4map}. A mapping extrapolated from that work is
shown in Figure~\ref{fig:zen4_mapping}.  For the remainder of this
paper we will assume Zen4 mappings as they are the most recent known
address mappings used in commercial products today.

Other works like Rubix~\cite{rubix} have suggested cryptographic-style address
mappings to help disperse activations evenly throughout the memory,
distributing column clusters across all rows in the DRAM. While this work reduces the number of activations that may accumulate on a single row, overall activation rates remain the same or are increased depending on which variation of Rubix is used.

\subsection{Rowhammer}
Rowhammer~\cite{rowhammer} is a well-studied attack vector in DRAM
which exploits electrical disturbance effects in DDR memories to
modify memory without direct access or privileges. These disturbance
effects occur between DRAM rows within an array, allowing for operations
upon one row to disturb nearby rows. Attackers induce high numbers of
activations on aggressor rows to accumulate these disturbances and
cause bit errors in nearby victim rows. The accumulated disturbance is
limited only by the occasional rewrites from DRAM refresh or accesses
to the victim row(s). Initially, in DDR3 technologies, failures would
require upwards of 100K nearby activations (referred to here as the
Rowhammer threshold tRH), but as newer DDR4 technologies were
introduced, this requirement dropped to as low as 4.8K
activations~\cite{rhrevisit}.  Modern DDR5 technologies incorporate
on-DRAM ECC and Rowhammer mitigations, making it difficult for the
vulnerability of these new technologies to be assessed. With lowering
tRH, the additional discovery of other potential disturbance
mechanisms such as Rowpress~\cite{rowpress}, and varying levels of industry adoption,
developing scalable low-overhead low-threshold mitigations has been an
uphill
battle.

\subsubsection{Academic Rowhammer Mitigations}

Many mitigation techniques have been proposed for Rowhammer \cite{mithril,graphene,randomized-rowswap,hydra,mint,aqua,archshield}. 
Initially, probabilistic methods like PARA~\cite{rowhammer} were deemed sufficient, but as tRH decreased
performance overheads were unsustainable. Many of these early mitigations were split between on-chip
(in-memory-controller) and off-chip (in-DRAM) designs
attempting to resolve this problem, some excluding the possibility of
collaboration between processor and DRAM manufacturers. Various
techniques were proposed, including denial-of-service \cite{blockhammer}, row
migration \cite{randomized-rowswap}, and tracking with selective refresh \cite{hydra,mithril,mint,graphene}. 
In anticipation of further decreases in tRH,
scalability has been the focus behind recent work, reducing overheads
sufficiently such that they are bearable into the far future. With the
onset of DDR5, new memory standards emerged, processor-DRAM
cooperation was standardized, and the field of Rowhammer mitigations
solidified.

\subsubsection{Rowhammer Mitigations in DDR5}

\begin{figure}[]
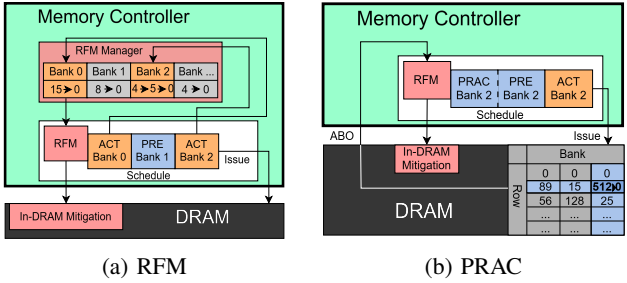

    \begin{subfigure}[]{0.48\columnwidth}
    \centering
    \includesvg[width=0.88\columnwidth]{figures/rfm_diagram.svg}
    \caption{RFM}\label{fig:rfm_diagram}
    \end{subfigure}
	\centering
    \begin{subfigure}[]{0.48\columnwidth}
    \centering
	\includesvg[width=0.95\columnwidth]{figures/prac_diagram.svg}\caption{PRAC}\label{fig:prac_diagram}
    \end{subfigure}
    \caption{RFM and PRAC Architecture}\label{fig:mitigation_diagrams}
\end{figure}

Support for two mitigation types have been added to the DDR5 JEDEC
standard, refresh management (RFM)~\cite{rfm} and Per Row Activation
Counting (PRAC)~\cite{prac1,prac2}.  Both of these components focus
primarily on off-chip mitigation and are differentiated by the amount
of state required for each and required design complexity. RFM, and
RFM-based mitigations like MINT~\cite{mint,mithril}, are considered low-cost
as they require little additional state, while PRAC and PRAC-like
mitigations~\cite{hydra,practical} require significant amounts of 
state in the form of counters in each DRAM row.  As depicted in
Figure~\ref{fig:rfm_diagram}, the refresh management command in RFM
allows the memory controller to reserve time for a vendor's on-DRAM
system to perform mitigation.  Due to the proprietary nature of
vendors' mitigation schemes and internal DRAM array layouts, the
memory controller blindly reserves time by issuing these commands
after a certain number of DRAM ACT commands, regardless of the actual
accumulated disturbances on DRAM rows or whether mitigations may be
invoked.  Per-row-activation counters (PRAC) work on a separate
principle as seen in Figure~\ref{fig:prac_diagram}, with per-row
counters in the DRAM allowing for exact tracking of how close rows are
to tolerated thresholds. While PRAC prevents unnecessary mitigation
from occurring (improving performance), the altered timings required to update the counters
incur significant energy overheads. Multiple works explore how these
techniques can be further improved, improving power and performance
overheads or rectifying security flaws \cite{autorfm,moat,when-mitigations-backfire,trespass,practical}.
RFM systems insert timing bubbles at a rate correlated to the vendor's defined RFM
threshold and rate of activations in the DRAM. As a vendor's defined
RFM threshold becomes lower (which results in a lower tolerated tRH),
or a workload induces a high ACT rate within the DRAM, RFM's overheads
become more costly. While PRAC's mitigation costs can be
hidden behind refresh (which can be insecure~\cite{moat}), the
significant overheads to PRE commands within the DRAM also cause slowdown,
though reportedly less than seen in simulation~\cite{Kim_2025}
when using friendly memory controller row-policy.  The exact
performance overheads to each of these schemes are variable, depending
on the workload and system configuration.  Within our simulated
system, PRAC (with a threshold of 512) and RFM (with a threshold of
16) had a 0.15\% and 5.1\% multi-core performance slowdown respectively. Both of
these overheads are tied to ACT commands, either entirely as in the
case of RFM or in part (PRE-ACT collisions in the memory controller) in the case
of PRAC.

If the absolute number of DRAM activations are increased
when running a given workload, it can be expected that the overhead (both in performance and energy) of whatever mitigation is in place would be significantly
increased. If underlying policies or other mechanisms are
introduced to decrease the number of DRAM activations, the expected
overheads would be significantly reduced. Likewise, one would expect that
techniques which can (in some cases dramatically) increase
activations, such as hardware prefetching, would likely have an impact
on overheads of these mitigation techniques.  We note that, despite
the large numbers of works in Rowhammer mitigations to date, we are
aware of no prior work to examine the intersection of hardware prefetching and
Rowhammer.  Further, we are not aware of any existing mitigation works that discuss using prefetching in the baseline system setup of
their experiments.

\begin{figure}[]
	\centering
    \includesvg[width=1.0\columnwidth]{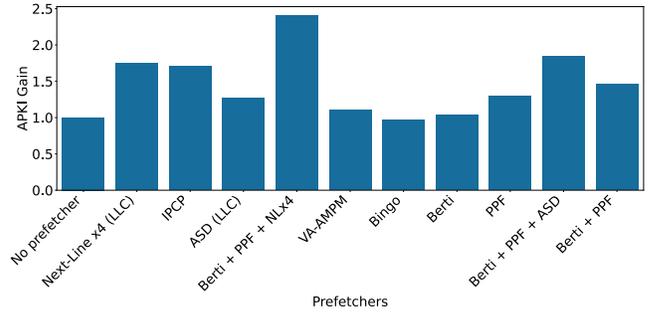}
        \caption{Geomean increase in activations per thousand instructions (APKI) when adding prefetchers on single-core workloads.}
        \label{fig:mit_overhead}
\end{figure}

\subsection{Hardware Prefetching}
\begin{figure}[]
	\centering
	\includesvg[width=.7\columnwidth]{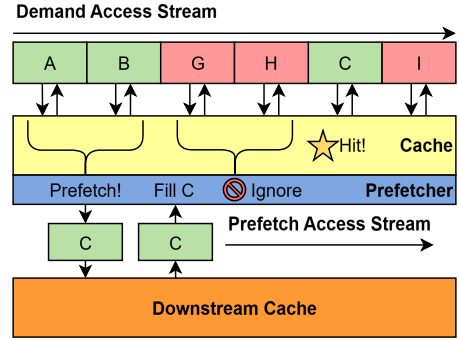}
        \caption{A simple hardware prefetcher}
        \label{fig:prefetcher_diagram}
\end{figure}

Hardware prefetchers speculatively fill CPU caches with
 timely and accurate data prior to their demand fetch. Simple prefetchers, such as shown in
Figure~\ref{fig:prefetcher_diagram}, use a static pattern
(fetching the next data block) while modern prefetchers attempt to identify and exploit complex
access patterns.

Most recent advanced prefetchers identify spatial (address-based) patterns and then use those patterns
to produce speculative future streams of accesses ahead of the current
demand~\cite{spp,ppf,berti,ipcp,ampm,bop}.  Some spatial  
prefetchers~\cite{spp}, are known to have been implemented in recent
commercial processors~\cite{exynos,Bruce_2023}.  Some prefetchers such as
Berti~\cite{berti} also incorporate temporal data to further improve
performance.

\subsubsection{Multi-level Hardware Prefetching}

While early prefetchers were typically designed in isolation, recent works examine prefetching in multiple
cache-levels. 
Modern prefetchers, including Berti \cite{berti}, PPF \cite{ppf},
and IPCP \cite{ipcp} each utilize some form of multi-level prefetching.
Since prefetches from one level can trigger prefetches from another
level, losses due to inaccuracy can become multiplicative and systemic
misbehaviors of one prefetcher trigger misbehaviors in
another. Likewise an LLC prefetcher, if it were added, would suffer
from the compounded misbehaviors of all instances of hardware
prefetchers within the system.

\subsubsection{Behaviors of Hardware Prefetchers}

As the amount of data pulled into caches by hardware prefetchers increases,
so does the strain on the memory system.  To understand why prefetchers
induce additional memory strain, one must consider \textit{why}
prefetchers improve performance.  Of the three types of cache misses,
prefetchers excel in their ability to prevent both compulsory and
capacity misses. The ability to issue prefetches ahead of time can
bring in fresh data or large data sets before they invoke cache
misses, and data can continue to be brought in until replacement
policies force the eviction of data within the current access
window. In this model, caches perform the role of a temporary buffer
for future data, rather than storage for frequently- or recently-used
data. Avoiding a capacity miss involves bringing in useful data while
also evicting useful data.

\begin{figure}[]
	\centering
	\includesvg[width=1.0\columnwidth]{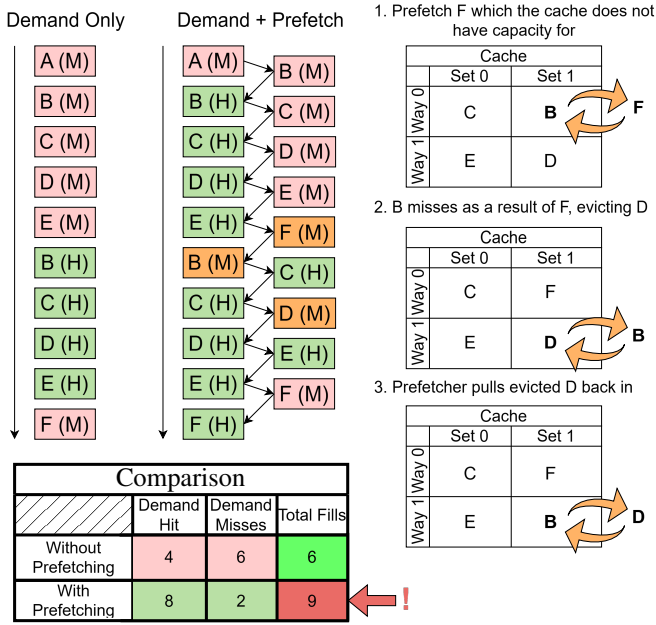}
        \caption{A next-line prefetcher induces cache churning}
        \label{fig:churn}
\end{figure}

Figure~\ref{fig:churn} depicts a repeating streaming access pattern
on a 4 entry LRU cache. Without prefetching, the stream has a low
cache hit rate of 40\%, but with a simple next-line prefetcher the hit rate
can be raised up to 80\%.  However, this comes at substantial cost to
the memory system. Blocks F, B, and D form an eviction chain that did not exist
in the original non-prefetching system. These additional misses lead to a 33\%
increase in cache fills, though they come at no apparent performance loss. 
The net effect is improved performance, but \textit{the cache is a cache in name only}, ultimately only serving as a
buffer for prefetched data prior to its use.  This is a problem
for prefetchers which exist in systems where memory bandwidth is
constrained as the increased memory access latency dominates
over the cache hit rate improvement.  This churning behavior spills
over into lower levels of the memory hierarchy, increasing activation rates in the
DRAM.
This is not isolated to a single workload or prefetcher, as shown in Figure~\ref{fig:mit_overhead}.

\subsection{Co-aware Prefetcher and Memory Controller Policies}
\begin{figure}[]
	\centering
	\includesvg[width=1.0\columnwidth]{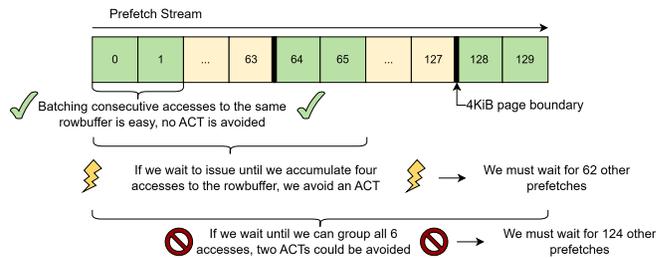}
        \caption{Batching more than one column cluster is impossible for existing methods}
        \label{fig:difficult_batching}
\end{figure}

While no techniques we are aware of have examined Rowhammer
mitigation in the context of a cache hierarchy with prefetching, some
techniques have shown that cooperation between memory controllers and
cache prefetchers can significantly increase rowbuffer hit rates.
DA-AMPM~\cite{daampm} and the Open-Page Minimalist~\cite{minpage}
scheduler both rely on batching prefetch requests to the same DRAM
rowbuffer into bursts, preventing small request reorderings from
inducing additional ACTs. These policies, which predate existing
Rowhammer mitigations, focus on short-term, reactive, activation avoidance.  Only DRAM requests that exist within the
system at the same time and going to the same place can be batched.
This limits which activations can be avoided and fails to avoid ACTs
to rowbuffers that may be avoidable but occur over longer timespans or
are part of larger access patterns. Importantly, prefetch streams that
cross over multiple column clusters cannot be batched retroactively.
This represents a substantial number of activations that cannot be
avoided.  For address mappings like that used by the Zen4 system~\cite{zen4map}, this
implies a pessimistic goal of 32 activations to
pull in the entire rowbuffer's content, under ideal operation.  This
is significantly more than the bare minimum of 1 activation that could
be required, but presents challenges.  As seen in
Figure~\ref{fig:difficult_batching}, larger addressable
memory sizes have resulted in the spatial distance between consecutive
column clusters increasing to the point where clusters within a rowbuffer 
are both spatially and temporally
isolated.  Waiting for these requests to coalesce to avoid an
activation effectively abandons those prefetches, delaying them by
thousands of cycles. 

Another difficulty comes from page size
restrictions; only a small number of cache lines are co-located within
a single rowbuffer for standard 4 KiB page sizes. This results in the
maximum batching size being limited to the number of column bits
addressable within the page offset (a single column cluster). Page sizes
are not fixed at 4 KiB, so systems which utilize larger page sizes can
circumvent this final issue. Even if all these problems are ignored, 
banks must handle requests serially, resulting in long access times and 
potential denial of the bank to others if too many column clusters are fetched at once.

\subsection{2 MiB Pages in Modern Systems}
Modern systems support several page sizes from KiBs to GiBs.  Using large pages reduces translation overheads, while the drawbacks of
larger page sizes can be avoided with proper operating system management \cite{vmware}.  OS kernels are very successful at automatically utilizing
these page sizes, so programs do not need to explicitly allocate large pages \cite{270443,thp,kaslr}.  Recent work has shown~\cite{trident,pageprefetch} that most if not all pages in modern systems running current OSes are 2 MiB.  Thus, unless otherwise specified, we assume 2 MiB pages for the remainder of this work.

\subsection{Summary}
Prior work in co-aware prefetcher and memory controller designs
are insufficient to exploit new technological advances and avoid performance
drawbacks from Rowhammer mitigations in modern memory.  Prefetcher designs have yet
to adapt to recent changes in DRAM technologies, and, depending on the workload,
may fail to meet the expected performance. Overcoming these obstacles is not trivial,
and will require a targeted prefetcher design.

\section{Design}
With consideration to the techniques, interactions, and obstacles described in Section \ref{sec:background}, we introduce the Optimized Row Access Prefetcher (ORAP).

\subsection{Overview}
\begin{figure}[]
	\centering
	\includesvg[width=1.0\columnwidth]{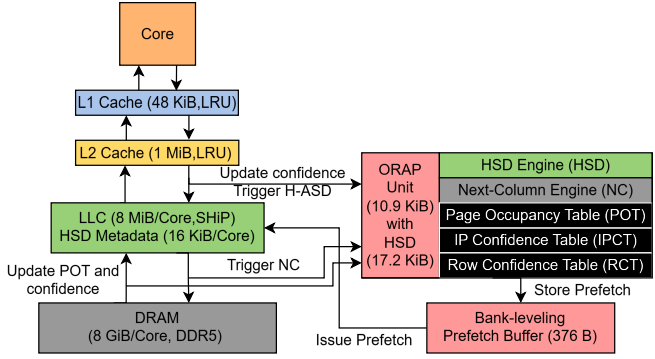}
        \caption{ORAP Architecture}
        \label{fig:orap}
\end{figure}
Depicted in Figure~\ref{fig:orap}, ORAP addresses the limitations of previous techniques via deep proactive prefetch streams. Instead of opportunistically reducing DRAM activations like DA-AMPM~\cite{daampm}, ORAP's primary goal is to actively avoid inducing activations. To achieve this, ORAP's Next-Column prefetch engine exploits long access streams to DRAM. Triggered on a cache miss, ORAP utilizes its instruction-pointer-confidence-table (IPCT) and row-confidence-table (RCT) to assess the likelihood that ORAP will be able to successfully cache other data within the missing request's target row. ORAP will then issue a variable number of prefetch requests, intending to piggy-back on the LLC miss to the DRAM and fetch additional information from the rowbuffer once it is opened. This results in large amounts of data being pulled into the LLC without additional activations. Under ideal circumstances, future DRAM activations are avoided and replaced with cache hits in the LLC.

A page-occupancy-table (POT) tracks the data recently brought into the cache to act as a filtering mechanism. To prevent the buildup of prefetch requests behind a DRAM bank, a bank-mapped prefetch buffer is maintained and managed (inspired by~\cite{blp_llc}). This queue serves to maximize BLP while also limiting the number of pending requests the prefetcher can have to a single bank.

Since Next-Column can only trigger on cache misses and targets a single bank, ORAP does not issue any mid-range prefetches by itself and does not exploit any bank parallelism. Therefore, the addition of a second prefetch engine to issue requests in a traditional fashion allows ORAP to have multiple banks serving Next-Column prefetches simultaneously. In this work, we utilize HSD (an ASD-inspired prefetcher \cite{asd} further discussed in Section~\ref{sec:HSD}) to fulfill this role.
For security purposes, ORAP's internal state is duplicated per-core and no information is shared between cores. 

Retaining many long-term prefetches in the LLC significantly disturbs the cache replacement policy. Some work has been done on integrating replacement policies with hardware prefetching \cite{kpc, pacman}.
We evaluate ORAP under SHiP~\cite{ship}, a popular academic policy that utilizes the program counter (IP) and the reuse distance of cache accesses when making replacement decisions. We find that prefetching from within the LLC greatly disturbs the standard SHiP replacement policy, resulting in performance loss. To rectify this, SHiP in ORAP's system ignores all prefetch accesses when collecting information or incrementing reuse counters. Unlike other works, this requires no additional state or logic in the SHiP policy.

\subsection{Avoiding DRAM Activations}
\begin{figure}[]
	\centering
	\includesvg[width=1.0\columnwidth]{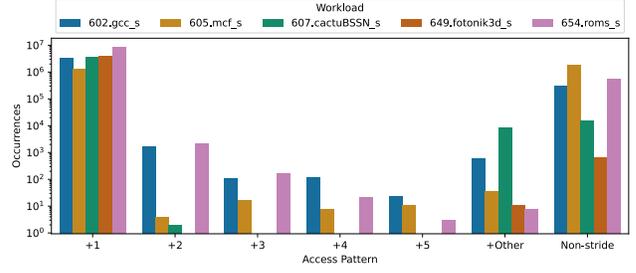}
    \caption{Occurrences of stride access patterns from LLC misses in SPEC2017 workloads}
    \label{fig:stride_pattern}
\end{figure}
\begin{figure}[]
	\centering
	\includesvg[width=1.0\columnwidth]{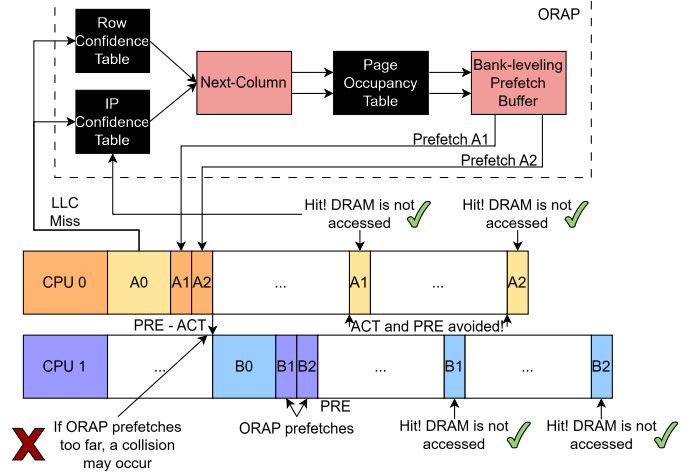}
    \caption{ORAP prefetches across columns}
    \label{fig:orap_column}
\end{figure}

Typical prefetchers are incapable of avoiding DRAM activations that are the result of column cluster crossings. The distance between subsequent column clusters is upwards of 64 blocks, which is outside the typical range that even aggressive prefetchers will attempt. Instead of pulling in all cache lines between the current and future column clusters, ORAP prefetches only the next column clusters. This reduces the required total prefetches by 33x at the cost of an irregular stride pattern. Fortunately, LLC misses are heavily biased towards long, streaming access patterns as shown in Figure~\ref{fig:stride_pattern}. This allows ORAP to use this irregular stride to pull in relevant data while retaining good usefulness. 

As illustrated in Figure~\ref{fig:orap_column} ORAP's Next-Column engine triggers only on cache misses. Depending on the confidence, ORAP will attempt to pull in a variable length of column clusters from the rowbuffer. Since ORAP issues prefetches on a cache miss, no additional activations are required to the target bank to serve these requests. Since these blocks now reside in the cache, future activations to access these column clusters are avoided. To prefetch within a single bank's rowbuffer, ORAP must know the logical-physical mapping used by the memory controller. This incurs a one-time setup cost to inform the prefetcher when the system initializes, and therefore has negligible performance impact.

\subsection{Confidence Challenges in Large Caches}
Prefetchers must choose how aggressively to issue prefetches. Confidence is a measure of the ability of a given prefetcher to issue useful prefetches and can be used as the metric to dynamically determine the aggression of the prefetcher. Confidence can be tracked globally or locally by associating confidence with some feature of accesses into the cache. 
Features include IPs such as in Fu et al. \cite{ipstride} but can also be spatial access patterns as in SPP \cite{spp} or latency in the case of Berti ~\cite{berti}. 
Utilizing confidence from feedback mechanisms can be difficult in large caches.  The feedback of whether a prefetch is useful or useless is temporally distant from when it was issued; the feedback is generally stale. Thus, the prefetcher throttles itself only after it has been behaving poorly for a long time. To account for this, ORAP treats confidence as a "pending prefetch counter".  Issuing a prefetch decrements the counter. Low confidence leads to few prefetches being issued before confidence drops to 0, while high confidence allows many prefetches to be issued before reaching 0. Confidence for a given IP and row is incremented once prefetches receive demand hits in the LLC. Confidence will slowly rise for prefetches whose usefulness allows a net gain in confidence, while slowly lower if usefulness falls below the necessary threshold.

\subsection{Hybrid Confidence}
\label{sec:hybrid_conf}

ORAP tracks confidence based on both the address space (DRAM row) and origin instruction (IP). 
Tracking only IP or row fails to exploit workloads which demonstrate only strong IP- or address-based patterns. By allowing IPs with high confidence to prefetch data in low confidence rows and IPs with poor confidence to prefetch in high confidence rows the coverage of ORAP is increased. 

The instruction-pointer-confidence-table (IPCT) and row-confidence-table (RCT) manage tracking IPs and rows using set-associative structures with LRU replacement. Each table is indexed by a 16-bit hash of the IP and row-id respectively. Since IP information for the original prefetch is not stored within the cache, cache hits resulting in a useful prefetch are attributed to the hitting IP, rather than the triggering IP of the prefetch. We find this sufficient since the IPCT is designed specifically for long single-IP streams which will likely be both the trigger and consumer of any prefetched data.

\begin{figure}[]
	\centering
	\includesvg[width=1.0\columnwidth]{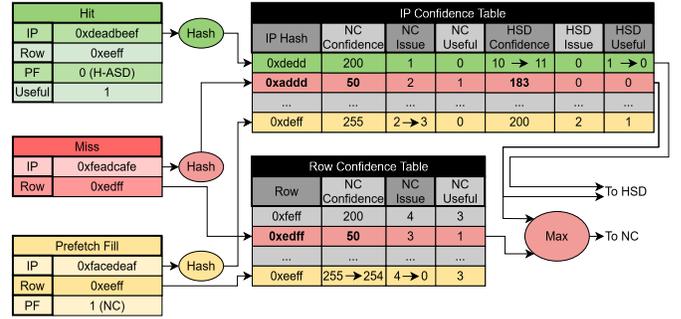}
        \caption{Management of IPCT and RCT Tables for Cache Operations}
        \label{fig:confidence_tables}
\end{figure}

The confidence tables are updated on different types of cache accesses, and the exact updates made are summarized in Figure~\ref{fig:confidence_tables}. For cache hits due to useful prefetches, the useful counter in the IPCT and RCT is incremented for the given row and IP, with only the IP used if the prefetch originated from HSD.
On a prefetch fill, the IPCT and RCT increment their issue counter similarly to the useful counter. If either counter rolls over, the confidence stored within the table is incremented if it was the useful counter or decremented if it was the issue counter. When confidence is required (a cache miss for Next-Column or any cache access for HSD), the confidence from the tables (only the IPCT in the case of HSD, or both in the case of Next-Column) is provided to each engine.

Depicted in Table~\ref{tab:prefetch_usefulness}, the values that the issue and useful counter maximums are set to impacts the target usefulness of ORAP. The issue counter maximum also provides absolute limits to the number of pending prefetches from a given IP or row that can reside in the cache without feedback (depending on how far the confidence is incremented). The potential range of pending prefetches are tabulated in Table~\ref{tab:pending_prefetches} for different confidence increments and issue counter maximums. Limiting the total prefetches a row or IP can have pending in the cache prevents long feedback times from resulting in many useless prefetches. It also remains flexible, such that if feedback times are reduced confidence is allowed to rise quickly and the confidence is not unnecessarily throttled.

Once the confidence value is obtained for a row or IP, these are translated into different metrics for use by the relevant prefetch engine. In the case of Next-Column, the confidence will determine how many column clusters into the future ORAP will attempt to prefetch. Because prefetching even one column cluster into the future is risky, ORAP will selectively prefetch on only a fraction of low-confidence triggers (9.5\% per 10 confidence, dropping no prefetches at 100 confidence). This also allows for triggers with 0 confidence to eventually be allowed to prefetch again, since they may slowly be allowed to issue prefetches and rebuild confidence.

A histogram probability is provided to HSD instead of a distance. This controls how deep HSD will prefetch, according to the probability of a stream reaching a certain depth. As confidence falls, the probability requirement increases and as confidence rises, the probability requirement falls.

\begin{table}[]
\resizebox{\columnwidth}{!}{
\centering
\begin{tabular}{
>{\columncolor[HTML]{343434}}c ccccc
>{\columncolor[HTML]{000000}}c 
>{\columncolor[HTML]{000000}}c }
\multicolumn{8}{c}{\cellcolor[HTML]{000000}{\color[HTML]{FFFFFF} Target Prefetch Usefulness}}                                                                                                                                                                                                                                                                                                                                                                                                                                   \\
\multicolumn{2}{c}{\cellcolor[HTML]{343434}{\color[HTML]{FFFFFF} }}                                                                                                               & \multicolumn{6}{c}{\cellcolor[HTML]{343434}{\color[HTML]{FFFFFF} Useful Counter Max}}                                                                                                                                                                                                                                                       \\
\multicolumn{2}{c}{\multirow{-2}{*}{\cellcolor[HTML]{343434}{\color[HTML]{FFFFFF} }}}                                                                                             & \cellcolor[HTML]{C0C0C0}{\color[HTML]{000000} 1}      & \cellcolor[HTML]{656565}{\color[HTML]{000000} 2}      & \cellcolor[HTML]{C0C0C0}{\color[HTML]{000000} 3}     & \cellcolor[HTML]{656565}{\color[HTML]{000000} 4}      & \cellcolor[HTML]{C0C0C0}{\color[HTML]{000000} 5}      & \cellcolor[HTML]{656565}{\color[HTML]{000000} 6}     \\
\cellcolor[HTML]{343434}{\color[HTML]{FFFFFF} }                                                                                & \cellcolor[HTML]{656565}{\color[HTML]{000000} 1} & \cellcolor[HTML]{FD6864}{\color[HTML]{000000} 100\%}  & \cellcolor[HTML]{000000}                              & \cellcolor[HTML]{000000}                             & \cellcolor[HTML]{000000}                              &                                                       &                                                      \\
\cellcolor[HTML]{343434}{\color[HTML]{FFFFFF} }                                                                                & \cellcolor[HTML]{C0C0C0}{\color[HTML]{000000} 2} & \cellcolor[HTML]{FFFE99}{\color[HTML]{000000} 50\%}   & \cellcolor[HTML]{FD6864}{\color[HTML]{000000} 100\%}  & \cellcolor[HTML]{000000}                             & \cellcolor[HTML]{000000}                              &                                                       &                                                      \\
\cellcolor[HTML]{343434}{\color[HTML]{FFFFFF} }                                                                                & \cellcolor[HTML]{656565}{\color[HTML]{000000} 3} & \cellcolor[HTML]{FFC499}{\color[HTML]{000000} 33\%}   & \cellcolor[HTML]{D7FF99}{\color[HTML]{000000} 66.7\%}   & \cellcolor[HTML]{FD6864}{\color[HTML]{000000} 100\%} & \cellcolor[HTML]{000000}                              &                                                       &                                                      \\
\cellcolor[HTML]{343434}{\color[HTML]{FFFFFF} }                                                                                & \cellcolor[HTML]{C0C0C0}{\color[HTML]{000000} 4} & \cellcolor[HTML]{FD6864}{\color[HTML]{000000} 25\%}   & \cellcolor[HTML]{FFFE99}{\color[HTML]{000000} 50\%}   & \cellcolor[HTML]{67FD9A}{\color[HTML]{000000} 75\%}  & \cellcolor[HTML]{FD6864}{\color[HTML]{000000} 100\%}  &                                                       &                                                      \\
\cellcolor[HTML]{343434}{\color[HTML]{FFFFFF} }                                                                                & \cellcolor[HTML]{656565}{\color[HTML]{000000} 5} & \cellcolor[HTML]{FD6864}{\color[HTML]{000000} 20\%}   & \cellcolor[HTML]{FFC499}{\color[HTML]{000000} 40\%}   & \cellcolor[HTML]{D7FF99}{\color[HTML]{000000} 60\%}  & \cellcolor[HTML]{67FD9A}{\color[HTML]{000000} 80\%}   & \cellcolor[HTML]{FD6864}{\color[HTML]{000000} 100\%}  &                                                      \\
\multirow{-6}{*}{\cellcolor[HTML]{343434}{\color[HTML]{FFFFFF} \begin{tabular}[c]{@{}c@{}}Issue\\ Counter\\ Max\end{tabular}}} & \cellcolor[HTML]{C0C0C0}{\color[HTML]{000000} 6} & \cellcolor[HTML]{FD6864}{\color[HTML]{000000} 16.7\%} & \cellcolor[HTML]{FFC499}{\color[HTML]{000000} 33.3\%} & \cellcolor[HTML]{FFFE99}{\color[HTML]{000000} 50\%}  & \cellcolor[HTML]{D7FF99}{\color[HTML]{000000} 66.7\%} & \cellcolor[HTML]{67FD9A}{\color[HTML]{000000} 83.3\%} & \cellcolor[HTML]{FD6864}{\color[HTML]{000000} 100\%}
\end{tabular}}
\caption{ORAP's targeted prefetch usefulness across different counter maximums.}
\label{tab:prefetch_usefulness}
\end{table}

\begin{table}[]
\centering
\begin{tabular}{
>{\columncolor[HTML]{343434}}c lllll}
\multicolumn{6}{c}{\cellcolor[HTML]{000000}{\color[HTML]{FFFFFF} Maximum Pending Prefetches}}                                                                                                                                                                                                                                                                                                            \\
\multicolumn{2}{l}{\cellcolor[HTML]{343434}{\color[HTML]{FFFFFF} }}                                                                                                                 & \multicolumn{4}{c}{\cellcolor[HTML]{343434}{\color[HTML]{FFFFFF} Confidence Increment}}                                                                                                                            \\
\multicolumn{2}{l}{\multirow{-2}{*}{\cellcolor[HTML]{343434}{\color[HTML]{FFFFFF} }}}                                                                                               & \cellcolor[HTML]{C0C0C0}{\color[HTML]{000000} 1}    & \cellcolor[HTML]{656565}{\color[HTML]{000000} 2}   & \cellcolor[HTML]{C0C0C0}{\color[HTML]{000000} 3}   & \cellcolor[HTML]{656565}{\color[HTML]{000000} 4}   \\
\cellcolor[HTML]{343434}{\color[HTML]{FFFFFF} }                                                                                  & \cellcolor[HTML]{656565}{\color[HTML]{000000} 1} & \cellcolor[HTML]{FFE3A1}{\color[HTML]{000000} 255}  & \cellcolor[HTML]{FFFFC7}{\color[HTML]{000000} 127} & \cellcolor[HTML]{FFFFC7}{\color[HTML]{000000} 85}  & \cellcolor[HTML]{FFFFFF}{\color[HTML]{000000} 63}  \\
\cellcolor[HTML]{343434}{\color[HTML]{FFFFFF} }                                                                                  & \cellcolor[HTML]{C0C0C0}{\color[HTML]{000000} 2} & \cellcolor[HTML]{FFB564}{\color[HTML]{000000} 510}  & \cellcolor[HTML]{FFE3A1}{\color[HTML]{000000} 255} & \cellcolor[HTML]{FFF3C7}{\color[HTML]{000000} 170} & \cellcolor[HTML]{FFFFC7}{\color[HTML]{000000} 127} \\
\cellcolor[HTML]{343434}{\color[HTML]{FFFFFF} }                                                                                  & \cellcolor[HTML]{656565}{\color[HTML]{000000} 3} & \cellcolor[HTML]{FD803C}{\color[HTML]{000000} 765}  & \cellcolor[HTML]{FFD676}{\color[HTML]{000000} 382} & \cellcolor[HTML]{FFE3A1}{\color[HTML]{000000} 255} & \cellcolor[HTML]{FFF3C7}{\color[HTML]{000000} 191} \\
\cellcolor[HTML]{343434}{\color[HTML]{FFFFFF} }                                                                                  & \cellcolor[HTML]{C0C0C0}{\color[HTML]{000000} 4} & \cellcolor[HTML]{FF6C1D}{\color[HTML]{000000} 1020} & \cellcolor[HTML]{FFB564}{\color[HTML]{000000} 510} & \cellcolor[HTML]{FFD676}{\color[HTML]{000000} 340} & \cellcolor[HTML]{FFE3A1}{\color[HTML]{000000} 255} \\
\cellcolor[HTML]{343434}{\color[HTML]{FFFFFF} }                                                                                  & \cellcolor[HTML]{656565}{\color[HTML]{000000} 5} & \cellcolor[HTML]{FF5B03}{\color[HTML]{000000} 1275} & \cellcolor[HTML]{FD803C}{\color[HTML]{000000} 637} & \cellcolor[HTML]{FFB564}{\color[HTML]{000000} 425} & \cellcolor[HTML]{FFD676}{\color[HTML]{000000} 318} \\
\multirow{-6}{*}{\cellcolor[HTML]{343434}{\color[HTML]{FFFFFF} \begin{tabular}[c]{@{}c@{}}Issue \\ Counter \\ Max\end{tabular}}} & \cellcolor[HTML]{C0C0C0}{\color[HTML]{000000} 6} & \cellcolor[HTML]{FF3420}{\color[HTML]{000000} 1530} & \cellcolor[HTML]{FD803C}{\color[HTML]{000000} 765} & \cellcolor[HTML]{FFB564}{\color[HTML]{000000} 510} & \cellcolor[HTML]{FFD676}{\color[HTML]{000000} 382}
\end{tabular}
\caption{ORAP's maximum pending prefetches across counter configurations.}
\label{tab:pending_prefetches}
\end{table}

\subsection{HSD}
\label{sec:HSD}
ORAP's Next-Column prefetcher by itself does not prefetch forward in the logical space and is thus limited to the bank the original LLC miss is on. This results in issues, as while ORAP can prefetch many blocks from a single bank, it cannot exploit any bank parallelism which limits the amount of DRAM bandwidth ORAP can utilize. We utilize a second prefetcher to learn LLC access patterns, predict future accesses, and cause cache misses to other banks that ORAP can then utilize. 

\begin{figure}[]
	\centering
	\includesvg[width=1.0\columnwidth]{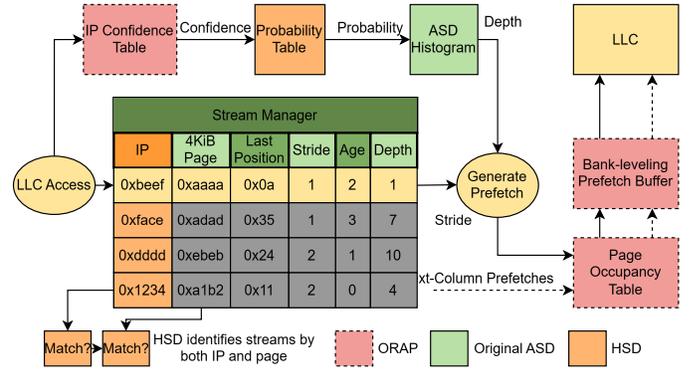}
        \caption{HSD's Architecture}
        \label{fig:asd_arch}
\end{figure}

ASD~\cite{asd} is a prefetcher designed with similar goals in mind, however, ASD was originally proposed in an era of much smaller DRAM systems and thus does not leverage new advancements. Originally, ASD was proposed as a LLC-miss prefetcher, residing within the memory controller which would prefetch blocks into a reserved prefetch buffer. This technique fails to utilize the space available in the CPU caches, and is limited to the information available within the memory controller. Instead we propose a new prefetcher, Hybrid Stream Detection (HSD), inspired by ASD but designed to modern constraints. HSD utilizes ASD's stream histogram technique to probabilistically issue prefetches in the logical space up to a depth of 64 blocks.  Unlike traditional ASD, our HSD utilizes a hybrid IP and address technique similar to what is described in Section \ref{sec:hybrid_conf} to identify streams. Shown in Figure~\ref{fig:asd_arch}, streams are collated either by matching IP or if they fall within the same 4 KiB page which yields better performance than using either method alone. Additionally, instead of prefetching forward with a minimum probability of 50\%, HSD utilizes the IPCT table to dynamically set the probability according to confidence. HSD's prefetch engine triggers on all LLC accesses and its generated prefetches are first sent through the LLC before heading to DRAM. The column prefetch engine is therefore able to exploit HSD's cache misses in the same way as the standard LLC misses. Running two prefetch engines within the same cache leads to many redundant prefetches being generated. To help synchronize information between ORAP's internal prefetch engines, the page-occupancy-table (POT) is used to filter both engines. 

\subsection{Preserving Bank-level-parallelism}
\begin{figure}[]
	\centering
    \includesvg[width=1.0\columnwidth]{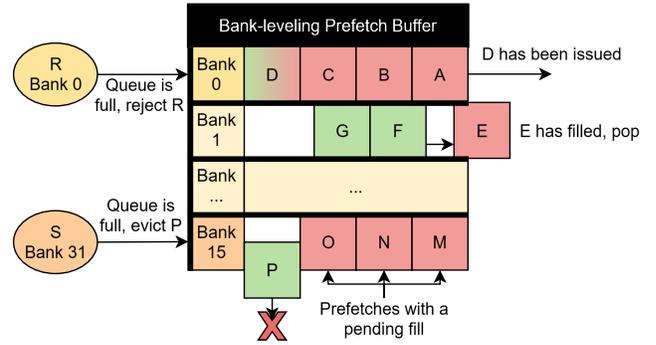}
        \caption{Bank-leveling Buffer Architecture}
        \label{fig:blp_arch}
\end{figure}
Issuing many prefetches to individual DRAM banks leads to undesirable impacts on DRAM bandwidth utilization. Consecutive requests to a single DRAM bank are required to wait serially to be served. If a demand request to the same bank arrives, it must either wait or the memory controller scheduler must preempt the previous prefetch requests, suffering two additional row activations but serving the critical demand request immediately. There are no additional benefits of issuing multiple requests simultaneously to the same DRAM bank from the LLC, and it only serves to occupy valuable MSHR space. Previous work~\cite{blp_llc} has tried to manage this through the LLC's allocation of MSHRs and scheduling of its internal miss queue. This is difficult to implement for modern systems, as the number of DRAM banks on a single chip have scaled from 8 to 32. Considering that these banks are duplicated across DRAM ranks and channels, it is not feasible to have queues for each physical DRAM bank. Additionally, not all memory configurations will utilize the maximum number of DRAM banks, leaving significant amounts of hardware inactive. 

Instead, pictured in Figure~\ref{fig:blp_arch}, we schedule prefetches according to DRAM bank as they leave ORAP rather than trying to reorder them as they miss in the cache. A set of bank-indexed buffers are shared between each instance of ORAP. As prefetches are generated, they are inserted into the buffer according to their destination bank ID. Unlike a standard queue, the prefetch buffer operates as an MSHR, holding buffer positions as occupied until prefetches return from issue or are dropped. If contention on a bank is high, the return time of prefetches to ORAP will throttle its requests to the bank due to the buffer filling. ORAP issues as many as two prefetches (one from two bank buffers) every cycle. To limit the amount of necessary hardware, the prefetch buffer is aliased into a smaller number of sub-buffers. This allows for efficient support for many different numbers of DRAM banks without excessive hardware requirements or large state left unused.

\section{Evaluation}
In this section, we evaluate ORAP against the current state-of-the-art hardware prefetcher configuration in single-core and 8-core simulations.

\subsection{Methodology}
We evaluate ORAP and other prefetchers within ChampSim, a micro-architectural trace-based simulator \cite{champsim}. To ensure conformity with DDR5 standards, ramulator2 \cite{ramulator2} is used as the DRAM model instead of ChampSim's internal model.

\begin{table}[]
\centering
\resizebox{\columnwidth}{!}{%
\begin{tabular}{clllll}
\rowcolor[HTML]{000000} 
\multicolumn{6}{c}{\cellcolor[HTML]{000000}{\color[HTML]{FFFFFF} \textbf{Cache   Configuration}}}                                                                                                                                                                                                                                                                                                                                                                                      \\                                                                                                                                                                                                                                                                                                                                                                                    
\rowcolor[HTML]{D0D0D0} 
\cellcolor[HTML]{000000}{\color[HTML]{000000} }                            & \multicolumn{1}{c}{\cellcolor[HTML]{D0D0D0}{\color[HTML]{000000} \textbf{Entries (Size)}}} & {\color[HTML]{000000} \textbf{Ways}} & {\color[HTML]{000000} \textbf{Latency (cycles)}}                                            & \multicolumn{2}{c}{\cellcolor[HTML]{D0D0D0}{\color[HTML]{000000} \textbf{MSHR Entries}}}                                                                                \\
\rowcolor[HTML]{A6A6A6} 
{\color[HTML]{000000} \textbf{I-TLB}}                                      & {\color[HTML]{000000} 256}                                                                 & {\color[HTML]{000000} 8}             & {\color[HTML]{000000} 1}                                                                    & \multicolumn{2}{l}{\cellcolor[HTML]{A6A6A6}{\color[HTML]{000000} 16}}                                                                                                   \\
\rowcolor[HTML]{D0D0D0} 
{\color[HTML]{000000} \textbf{D-TLB}}                                      & {\color[HTML]{000000} 96}                                                                  & {\color[HTML]{000000} 6}             & {\color[HTML]{000000} 1}                                                                    & \multicolumn{2}{l}{\cellcolor[HTML]{D0D0D0}{\color[HTML]{000000} 16}}                                                                                                   \\
\rowcolor[HTML]{A6A6A6} 
{\color[HTML]{000000} \textbf{S-TLB}}                                      & {\color[HTML]{000000} 2048}                                                                & {\color[HTML]{000000} 16}            & {\color[HTML]{000000} 7}                                                                    & \multicolumn{2}{l}{\cellcolor[HTML]{A6A6A6}{\color[HTML]{000000} 8}}                                                                                                    \\
\rowcolor[HTML]{D0D0D0} 
{\color[HTML]{000000} \textbf{L1I}}                                        & {\color[HTML]{000000} 32 KiB}                                                              & {\color[HTML]{000000} 8}             & {\color[HTML]{000000} 4}                                                                    & \multicolumn{2}{l}{\cellcolor[HTML]{D0D0D0}{\color[HTML]{000000} 16}}                                                                                                   \\
\rowcolor[HTML]{A6A6A6} 
{\color[HTML]{000000} \textbf{L1D}}                                        & {\color[HTML]{000000} 48 KiB}                                                              & {\color[HTML]{000000} 12}            & {\color[HTML]{000000} 5}                                                                    & \multicolumn{2}{l}{\cellcolor[HTML]{A6A6A6}{\color[HTML]{000000} 32}}                                                                                                   \\
\rowcolor[HTML]{D0D0D0} 
{\color[HTML]{000000} \textbf{L2}}                                         & {\color[HTML]{000000} 1 MiB}                                                               & {\color[HTML]{000000} 16}            & {\color[HTML]{000000} 10}                                                                   & \multicolumn{2}{l}{\cellcolor[HTML]{D0D0D0}{\color[HTML]{000000} 64}}                                                                                                   \\
\rowcolor[HTML]{A6A6A6} 
{\color[HTML]{000000} \textbf{LLC}}                                        & {\color[HTML]{000000} 8 MiB/Core}                                                          & {\color[HTML]{000000} 16}            & {\color[HTML]{000000} 60}                                                                   & \multicolumn{2}{l}{\cellcolor[HTML]{A6A6A6}{\color[HTML]{000000} 40/Core}}                                                                                              \\
\multicolumn{1}{l}{}                                                       &                                                                                            &                                      &                                                                                             &                                                                                      &                                                                                  \\
\multicolumn{2}{c}{\cellcolor[HTML]{000000}{\color[HTML]{FFFFFF} \textbf{Core Configuration}}}                                                                          &                                      & \multicolumn{3}{c}{\cellcolor[HTML]{000000}{\color[HTML]{FFFFFF} \textbf{Memory Controller Configuration}}}                                                                                                                                                           \\
\cellcolor[HTML]{D0D0D0}{\color[HTML]{000000} \textbf{Frequency}}          & \cellcolor[HTML]{D0D0D0}{\color[HTML]{000000} 4 GHz}                                       & {\color[HTML]{000000} }              & \multicolumn{1}{c}{\cellcolor[HTML]{D9D9D9}{\color[HTML]{000000} \textbf{Channels}}}        & \multicolumn{2}{l}{\cellcolor[HTML]{D9D9D9}{\color[HTML]{000000} 2 1-core, 4 8-core}}                                                                                   \\
\cellcolor[HTML]{A6A6A6}{\color[HTML]{000000} \textbf{Branch   Predictor}} & \cellcolor[HTML]{A6A6A6}{\color[HTML]{000000} Hashed Perceptron}                           & {\color[HTML]{000000} }              & \multicolumn{1}{c}{\cellcolor[HTML]{A6A6A6}{\color[HTML]{000000} \textbf{Ranks}}}           & \multicolumn{2}{l}{\cellcolor[HTML]{A6A6A6}{\color[HTML]{000000} 1 1-core, 2 8-core}}                                                                                   \\
\cellcolor[HTML]{D0D0D0}{\color[HTML]{000000} \textbf{Issue Width}}        & \cellcolor[HTML]{D0D0D0}{\color[HTML]{000000} 6}                                           & {\color[HTML]{000000} }              & \multicolumn{1}{c}{\cellcolor[HTML]{D9D9D9}{\color[HTML]{000000} \textbf{Scheduler}}}       & \multicolumn{2}{l}{\cellcolor[HTML]{D9D9D9}{\color[HTML]{000000} Minimalist Open Page}}                                                                                 \\
\cellcolor[HTML]{A6A6A6}{\color[HTML]{000000} \textbf{Retire Width}}       & \cellcolor[HTML]{A6A6A6}{\color[HTML]{000000} 8}                                           & {\color[HTML]{000000} }              & \multicolumn{1}{c}{\cellcolor[HTML]{A6A6A6}{\color[HTML]{000000} \textbf{Row Policy}}}      & \multicolumn{2}{l}{\cellcolor[HTML]{A6A6A6}{\color[HTML]{000000} Adaptive Row Policy}}                                                                                  \\
\cellcolor[HTML]{D0D0D0}{\color[HTML]{000000} \textbf{Scheduler Size}}     & \cellcolor[HTML]{D0D0D0}{\color[HTML]{000000} 205}                                         & {\color[HTML]{000000} }              & \multicolumn{1}{c}{\cellcolor[HTML]{D9D9D9}{\color[HTML]{000000} \textbf{Address Mapping}}} & \multicolumn{2}{l}{\cellcolor[HTML]{D9D9D9}{\color[HTML]{000000} Zen4 (8 GiB, 64 GiB)}}                                                                                 \\
\cellcolor[HTML]{A6A6A6}{\color[HTML]{000000} \textbf{ROB Size}}           & \cellcolor[HTML]{A6A6A6}{\color[HTML]{000000} 512}                                         & {\color[HTML]{000000} }              & \multicolumn{1}{c}{\cellcolor[HTML]{A6A6A6}{\color[HTML]{000000} \textbf{RFM Threshold}}}   & \multicolumn{2}{l}{\cellcolor[HTML]{A6A6A6}{\color[HTML]{000000} 16}}                                                                                                   \\
\cellcolor[HTML]{D0D0D0}{\color[HTML]{000000} \textbf{LQ Size}}            & \cellcolor[HTML]{D0D0D0}{\color[HTML]{000000} 192}                                         & {\color[HTML]{000000} }              & \multicolumn{1}{c}{\cellcolor[HTML]{D9D9D9}{\color[HTML]{000000} \textbf{PRAC Threshold}}}  & \multicolumn{2}{l}{\cellcolor[HTML]{D9D9D9}{\color[HTML]{000000} 512}}                                                                                                  \\
\cellcolor[HTML]{A6A6A6}{\color[HTML]{000000} \textbf{SQ Size}}            & \cellcolor[HTML]{A6A6A6}{\color[HTML]{000000} 114}                                         & {\color[HTML]{000000} }              & \multicolumn{1}{c}{\cellcolor[HTML]{A6A6A6}{\color[HTML]{000000} \textbf{Blast Radius}}}    & \multicolumn{2}{l}{\cellcolor[HTML]{A6A6A6}{\color[HTML]{000000} 2}}                                                                                                    \\
\multicolumn{1}{l}{}                                                       &                                                                                            &                                      &                                                                                             &                                                                                      &                                                                                  \\
\multicolumn{2}{c}{\cellcolor[HTML]{000000}{\color[HTML]{FFFFFF} DRAM Configuration}}                                                                                   &                                      & \multicolumn{3}{c}{\cellcolor[HTML]{000000}{\color[HTML]{FFFFFF} \textbf{DRAM Timings}}}                                                                                                                                                                              \\
\cellcolor[HTML]{A6A6A6}{\color[HTML]{000000} \textbf{Standard}}           & \cellcolor[HTML]{A6A6A6}{\color[HTML]{000000} DDR5}                                        &                                      & \cellcolor[HTML]{000000}                                                                    & \multicolumn{1}{r}{\cellcolor[HTML]{A6A6A6}{\color[HTML]{000000} \textbf{Standard}}} & \multicolumn{1}{r}{\cellcolor[HTML]{A6A6A6}{\color[HTML]{000000} \textbf{PRAC}}} \\
\cellcolor[HTML]{D0D0D0}{\color[HTML]{000000} \textbf{MTPS}}               & \cellcolor[HTML]{D0D0D0}{\color[HTML]{000000} 6400}                                        &                                      & \cellcolor[HTML]{D0D0D0}{\color[HTML]{000000} \textbf{nCL}}                                 & \cellcolor[HTML]{D0D0D0}{\color[HTML]{000000} 16.25 ns}                                    & \cellcolor[HTML]{D0D0D0}{\color[HTML]{000000} 16.25 ns}                                \\
\cellcolor[HTML]{A6A6A6}{\color[HTML]{000000} \textbf{Rows}}               & \cellcolor[HTML]{A6A6A6}{\color[HTML]{000000} 65536}                                       &                                      & \cellcolor[HTML]{A6A6A6}{\color[HTML]{000000} \textbf{nRCD}}                                & \cellcolor[HTML]{A6A6A6}{\color[HTML]{000000} 16.25 ns}                                    & \cellcolor[HTML]{A6A6A6}{\color[HTML]{000000} 16.25 ns}                                \\
\cellcolor[HTML]{D0D0D0}{\color[HTML]{000000} \textbf{Columns}}            & \cellcolor[HTML]{D0D0D0}{\color[HTML]{000000} 1024}                                        &                                      & \cellcolor[HTML]{D0D0D0}{\color[HTML]{000000} \textbf{nRP}}                                 & \cellcolor[HTML]{D0D0D0}{\color[HTML]{000000} 16.25 ns}                                    & \cellcolor[HTML]{D0D0D0}{\color[HTML]{000000} 36.25 ns}                               \\
\cellcolor[HTML]{A6A6A6}{\color[HTML]{000000} \textbf{Banks}}              & \cellcolor[HTML]{A6A6A6}{\color[HTML]{000000} 4}                                           &                                      & \cellcolor[HTML]{A6A6A6}{\color[HTML]{000000} \textbf{nRAS}}                                & \cellcolor[HTML]{A6A6A6}{\color[HTML]{000000} 32.5 ns}                                   & \cellcolor[HTML]{A6A6A6}{\color[HTML]{000000} 16.25 ns}                                \\
\cellcolor[HTML]{D0D0D0}{\color[HTML]{000000} \textbf{Bankgroups}}         & \cellcolor[HTML]{D0D0D0}{\color[HTML]{000000} 4 1-core, 8 8-core}                          &                                      & \cellcolor[HTML]{D0D0D0}{\color[HTML]{000000} \textbf{nRC}}                                 & \cellcolor[HTML]{D0D0D0}{\color[HTML]{000000} 48 ns}                                   & \cellcolor[HTML]{D0D0D0}{\color[HTML]{000000} 52 ns}                               \\
\cellcolor[HTML]{A6A6A6}{\color[HTML]{000000} \textbf{Density}}            & \cellcolor[HTML]{A6A6A6}{\color[HTML]{000000} 16 Gib}                                      &                                      & \cellcolor[HTML]{A6A6A6}{\color[HTML]{000000} \textbf{nWR}}                                 & \cellcolor[HTML]{A6A6A6}{\color[HTML]{000000} 30 ns}                                    & \cellcolor[HTML]{A6A6A6}{\color[HTML]{000000} 10 ns}                                \\
\cellcolor[HTML]{D0D0D0}{\color[HTML]{000000} \textbf{Channel Width}}      & \cellcolor[HTML]{D0D0D0}{\color[HTML]{000000} 32-bit}                                      &                                      & \cellcolor[HTML]{D0D0D0}{\color[HTML]{000000} \textbf{nRTP}}                                & \cellcolor[HTML]{D0D0D0}{\color[HTML]{000000} 7.5 ns}                                    & \cellcolor[HTML]{D0D0D0}{\color[HTML]{000000} 5 ns}                                \\
\cellcolor[HTML]{A6A6A6}{\color[HTML]{000000} \textbf{Physical Size}}      & \cellcolor[HTML]{A6A6A6}{\color[HTML]{000000} 8 GiB/Core}                                  &                                      &                                                                                             &                                                                                      &                                                                                  \\
\cellcolor[HTML]{D0D0D0}{\color[HTML]{000000} \textbf{Refresh Period}}     & \cellcolor[HTML]{D0D0D0}{\color[HTML]{000000} 32 ms}                                       &                                      &                                                                                             &                                                                                      &                                                                                 
\end{tabular}%
}
\caption{System Configuration}
\label{tab:system_config}
\end{table}

We use the SimPoint methodology~\cite{simpoint} on traces from the SPEC2006, SPEC2017, GAP, and XSBench benchmark suites \cite{spec2006,spec2017,gaps,xsbench}.
For our single-core simulations traces are first run for 200M instructions before data collection, followed with data collection for 200M instructions. For 8-core simulations, the simulation is run until all 8-cores have executed 200M instructions, and then statistics are collected until they all have executed another 200M instructions. To support ORAP, ChampSim's page sizes are set to 2 MiB which is supported on most modern machines \cite{thp,arm_pages,intel_pages,270443,amd_npages,amd_pages}.

We compare ORAP against various academic prefetchers in single-core simulations, including ASD, VA-AMPM, Bingo, Berti, IPCP, and SPP-PPF. In multi-core, ORAP is evaluated against the state-of-the-art configuration of Berti in the L1D and SPP-PPF in the L2C. We evaluate under both PRAC and RFM mitigations. The reverse-engineered mappings from AMD's Zen4 processor are used for DRAM mappings within the memory controller, as they represent the most-recent example of industry mappings \cite{zen4map}. For further accuracy, ChampSim is modified to support prefetch promotion, a performance optimization where demands can mark prefetches they are waiting on for prioritization by the memory system. Additionally, ramulator2 utilizes an adaptive row policy and scheduler from prior work \cite{Kim_2025,minpage}. Our system configuration is listed in Table~\ref{tab:system_config}, which is modeled after recent Zen4 cores, including their 64 MiB 3D-integrated caches~\cite{zen5}.

For evaluating energy results, we report normalized DRAM dynamic energy per instruction (DEPI) which serves as a unit of the energy efficiency of the system. Energies are obtained through ramulator2's energy model. Under an Adaptive Row Policy, we identify dynamic energy as energy expended over an idle system, including the energy to execute DRAM commands and keep banks active.

We evaluate ORAP on all workloads in the selected benchmark suites, with the exception of bc.kron, bc.urand, bfs.urand, sssp.urand, sssp.kron, sssp.twitter, sssp.web, and tc.urand from GAP which were excluded due to simulation infrastructure constraints. Every simpoint for each workload was simulated in single-core. For 8-core simulations, 73 mixes were randomly compiled from all simpoints.

\begin{table}[]
\resizebox{\columnwidth}{!}{%
\begin{tabular}{|l|l|l|}
\hline
\rowcolor[HTML]{000000} 
{\color[HTML]{FFFFFF} \textbf{Structure}}    & {\color[HTML]{FFFFFF} \textbf{Breakdown}}                                                                                                                                                                                       & {\color[HTML]{FFFFFF} \textbf{State}}                              \\ \hline
\rowcolor[HTML]{D9D9D9} 
IP Confidence Table                          & \begin{tabular}[c]{@{}l@{}}1024 entries, 8   ways, 25 (38) bits per entry:\\      9-bit IP tag, 3-bit LRU, 8-bit confidence / engine, \\      3-bit issue counter (max 5) / engine, 2-bit useful counter (max 4) / engine\end{tabular} & \begin{tabular}[c]{@{}l@{}}3.13 KiB\\      (4.75 KiB)\end{tabular} \\ \hline
\rowcolor[HTML]{A6A6A6} 
Row Confidence   Table                       & \begin{tabular}[c]{@{}l@{}}1024   entries, 8 ways, 25 bits per entry:\\      9-bit row tag, 3-bit LRU, 8-bit confidence, 3-bit issue counter (max 5),\\      2-bit useful counter (max 4)\end{tabular}                                 & 3.13 KiB                                                              \\ \hline
\rowcolor[HTML]{D9D9D9} 
Page Occupancy   Table                       & \begin{tabular}[c]{@{}l@{}}256   entries, 8 ways, 100 bits per entry:\\      36-bit 4 KiB page id, 64-bit page occupancy map\end{tabular}                                                                                       & 3.13 KiB                                                          \\ \hline
\rowcolor[HTML]{A6A6A6} 
Bank-leveling   Buffer                       & \begin{tabular}[c]{@{}l@{}}64 total entries / core (16 sub-buffers), 47 bits per entry:\\      42-bit block number, 3-bit cpu id, 1-bit engine-id,\\      1-bit pending flag\end{tabular}                                                            & 376 B                                                              \\ \hline
\rowcolor[HTML]{D9D9D9} 
HSD   Histogram                            & 2   histograms, 64 entries each, 13 bits per entry: access count                                                                                                                                                                & (1.63  KiB)                                                       \\ \hline
\rowcolor[HTML]{A6A6A6} 
HSD State   Machine                        & 3-bit   state machine register, 13-bit epoch counter                                                                                                                                                                            & (2 B)                                                              \\ \hline
\rowcolor[HTML]{D9D9D9} 
HSD Stream   Manager                       & \begin{tabular}[c]{@{}l@{}}32   streams, 106 bits per stream:\\      48-bit IP, 36-bit 4 KiB page id, 6-bit last-offset, \\      5-bit depth, 3-bit stride,  8-bit age\end{tabular}                                             & (424 B)                                                        \\ \hline
\rowcolor[HTML]{A6A6A6} 
LLC                                          & \begin{tabular}[c]{@{}l@{}}128   K lines, 16 KiB:\\      log2(engines)-bit prefetch-engine id per block\end{tabular}                                                                                                            & (16 KiB)                                                           \\ \hline
\rowcolor[HTML]{000000} 
{\color[HTML]{FFFFFF} \textbf{ORAP}}         & {\color[HTML]{FFFFFF} \textbf{}}                                                                                                                                                                                                & {\color[HTML]{FFFFFF} \textbf{9.76 KiB}}                         \\ \hline
\rowcolor[HTML]{000000} 
{\color[HTML]{FFFFFF} \textbf{ORAP + HSD}} & {\color[HTML]{FFFFFF} \textbf{}}                                                                                                                                                                                                & {\color[HTML]{FFFFFF} \textbf{29.42 KiB}}                          \\ \hline
\end{tabular}%

}
\caption{State Overhead of ORAP}
\label{tab:state_orap}
\end{table}

\subsection{Results}
\subsubsection{Storage Overheads}
Compared to the state-of-the-art Berti and PPF prefetching scheme, Berti + ORAP and Berti + ORAP + HSD reduce necessary state by 67.7\% (9.76 KiB) and 22.7\% (30.66 KiB) respectively. As shown in Table~\ref{tab:state_orap}, ORAP requires relatively little state per core resulting in a small area overhead to the entire LLC (0.37\%). The primary overhead of adding HSD comes from differentiating Next-Column and HSD prefetches, which requires an additional bit in each line of the cache.

\subsubsection{Multi-core Performance}\label{sec:multicore_performance}
\begin{figure}[]
	\centering
	\includesvg[width=1.0\columnwidth]{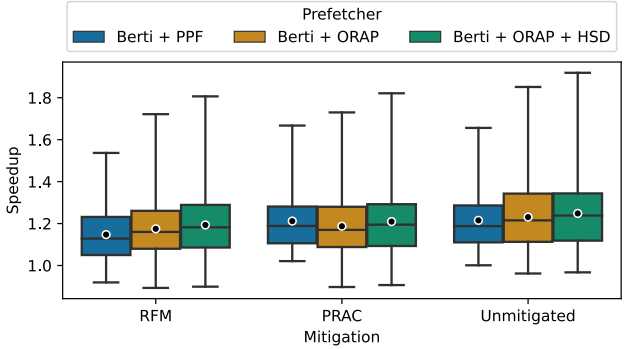}
        \caption{Average multi-core speedup under different mitigations}
        \label{fig:multicore_speedup}
\end{figure}

\begin{figure}[]
	\centering
	\includesvg[width=1.0\columnwidth]{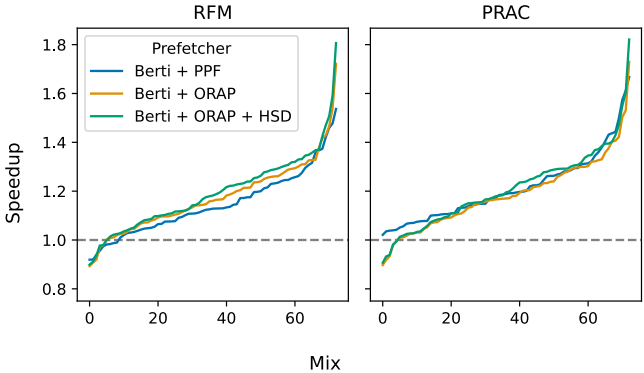}
        \caption{Sorted speedups of mixes under RFM and PRAC mitigations}
        \label{fig:speedup_lineplot}
\end{figure}

Multi-core speedup across all 73 mixes is provided in Figure~\ref{fig:multicore_speedup}, normalized to a non-prefetching system.  The average mix speedup over Berti + PPF for Berti + ORAP + HSD was 4.6\% in RFM systems, 3.3\% speedup in unmitigated systems, and a 0.24\% slowdown in PRAC systems. ORAP has both higher maximum and lower minimum speedups (indicated by the whiskers) than Berti + PPF. From Figure~\ref{fig:speedup_lineplot}, it can be seen that ORAP is consistently better than Berti + PPF under RFM, yet struggles on the lowest-performing workloads under PRAC. As we will see, however, while ORAP performance under PRAC is slightly worse, it significantly reduces activation rates and energy consumption.

Berti + PPF saw a  6.8\% and 0.35\% slowdown for RFM and PRAC respectively, compared to the non-prefetching overheads of 5.1\% and 0.15\%. Interestingly, ORAP suffers a 5.48\% and 3.88\% slowdown over its unmitigated counterpart for RFM and PRAC. This suggests that while ORAP reduces the overhead of RFM mitigations, it is sensitive to the timing changes under PRAC. PRAC increases precharge times over baseline, meaning that rowbuffer conflicts come at greater penalties. Since ORAP increases the number of blocks fetched from an opened rowbuffer, the Adaptive Row Policy in the memory controller may learn to hold rowbuffers open longer, increasing the risk of conflicts and the total PRAC overhead. The recent work by Kim et al. \cite{Kim_2025} explores these PRAC interactions with row policies in more detail.
\begin{figure*}[]
	\centering
	\includesvg[width=2.0\columnwidth]{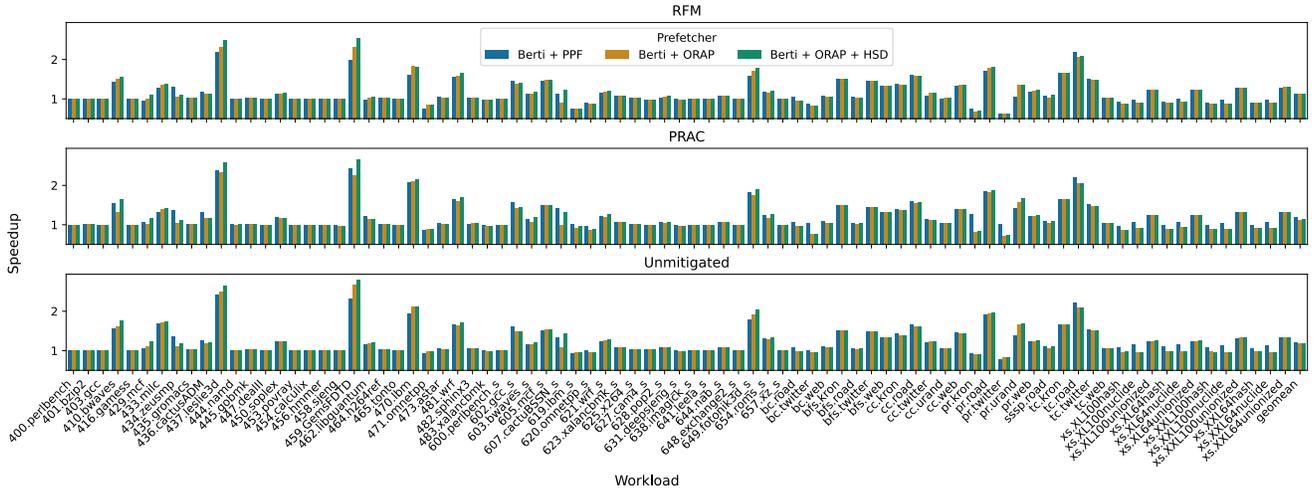}
        \caption{Single-core speedup of ORAP vs. Berti + PPF}
        \label{fig:singlecore_speedup}
\end{figure*}

\subsubsection{DRAM Activation Rates and Dynamic Energy}
\begin{figure}[]
\centering

	\includesvg[width=1.0\columnwidth]{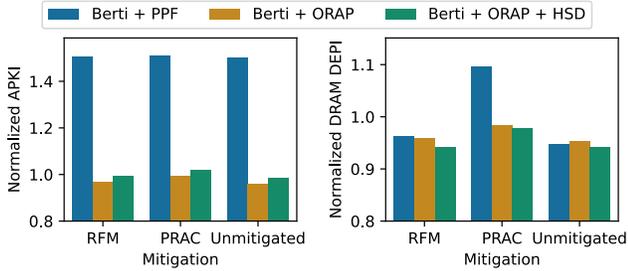}
        \caption{DRAM activation rates (left) compared to DRAM dynamic energy (right)}
        \label{fig:multicore_apki}
\end{figure}

Figure~\ref{fig:multicore_apki} shows the multi-core activation rates (activations per thousand instructions) of Berti + PPF and ORAP normalized to a non-prefetching system.
Compared to Berti + PPF's activation rates, rates are reduced by 54.2\% and 51.3\% for Berti + ORAP and Berti + ORAP + HSD respectively.
This reduction was the primary goal of ORAP's design, but fails to achieve the loftier goals of reducing rates to below baseline. This suggests that further reduction in rates would require
redesigns within ORAP or expanding the purview of the design into memory controller policy and address mappings.

We analyze the dynamic energy of the DRAM normalized to the baseline energy overhead of each mitigation.
Generally, we find that dynamic energy is reduced over a non-prefetching system, with Berti + PPF showing a 5.34\% reduction over baseline for an unmitigated system.
For Berti + ORAP + HSD, energy is further reduced by 2.02\% and 0.55\% for RFM and unmitigated systems. The difference in dynamic energy is most drastic under
PRAC mitigations. Since the rates of DRAM activations are substantially higher under Berti + PPF, so are the energy overheads associated with precharges. 
ORAP's reduction in activation rates translates to significant energy reduction: 11.8\%.

\subsubsection{Prefetch Usefulness and DRAM Bandwidth}
ORAP's prefetching strategy is aggressive, and while it tries to preserve the usefulness of prefetches, it is designed to prefetch in a way that is counter to that goal. ORAP prefetches deeply into pages,
so preserving usefulness is difficult.
We find that ORAP as a replacement for PPF has lower overall usefulness, dropping from 57\% to 47\%. ORAP prefetches in the LLC, so space is abundant and small drops in usefulness are less impactful than if it were placing in the L2. ORAP avoids DRAM activations when issuing prefetches, so these prefetches consume relatively little energy and are not typically disruptive.

\begin{figure}[]
	\centering
	\includesvg[width=1.0\columnwidth]{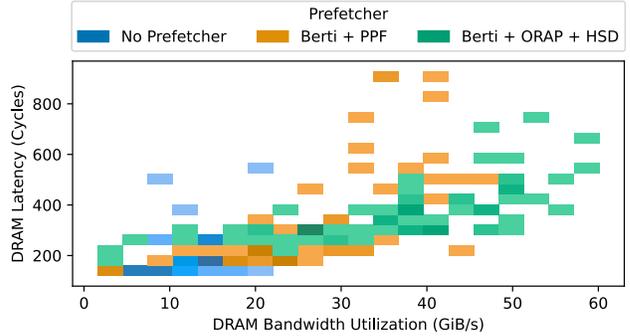}
        \caption{Bandwidth consumption between ORAP and Berti + PPF under an RFM-mitigated system}
        \label{fig:bandwidth}
\end{figure}

Compared to Berti + PPF, average bandwidth utilization rises from 31.4 GiB/s to 39.6 GiB/s across our mixes. At the same time, average DRAM latency rose from 281 cycles to 323 cycles.
Figure~\ref{fig:bandwidth} displays the distribution of different mixes across DRAM bandwidth and latency for Berti + PPF and Berti + ORAP + HSD.
Note that for Berti + PPF as bandwidth utilization begins to increase DRAM latency rises sharply. However, Berti + ORAP + HSD's latency rises gradually, and has a lower peak latency and higher peak
bandwidth utilization than Berti + PPF. We attribute this to two complementary negative impacts of Berti + PPF's prefetching behavior. First, we noted that increased rowbuffer conflicts could cause performance loss in Section~\ref{sec:multicore_performance}. Although Berti + PPF is unlikely to cause rowbuffers to remain open for long times like ORAP, it is likely to access rowbuffers which are not open to the requested row. As a result, scheduling policies within the memory controller may choose to delay these packets to serve packets that will result in rowbuffer hits, which in turn limits Berti + PPF's ability to exploit DRAM bandwidth. The second cause for this behavior is the reduced latency of rowbuffer hits. Since ORAP requests multiple blocks from a single open rowbuffer, its packets always hit, and as a result are not only served quicker, but have an overall lower access latency. These combined behaviors mean that ORAP can utilize more DRAM bandwidth while having less overall impact on the performance of the memory system, conditioned on timings that minimize the disruption of long contiguous streams of rowbuffer hits.

\subsubsection{Single-core Performance}
The single-core performance of various prefetchers under RFM and PRAC mitigations can be seen in Figure~\ref{fig:mitigation_effect_on_prefetchers}. We further breakdown their performance across workloads for Berti + PPF, Berti + ORAP, and Berti + ORAP + HSD in Figure~\ref{fig:singlecore_speedup}. Compared to a no-prefetching baseline, Berti + ORAP + HSD provides 12.8\% geomean speedup and outperforms the state-of-the-art by 1.2\% under RFM. In PRAC and unmitigated systems, ORAP incurs a 3.28\% and 0.4\% performance loss respectively. We find that ORAP outperforms Berti + PPF on many SPEC2006, SPEC2017, and GAP workloads. ORAP is consistently worse for many of the XSBench workloads, though this margin is significantly reduced under RFM mitigations.

\section{Conclusion And Future Work}
ORAP offers considerable improvements over the current state-of-the-art in prefetching. ORAP outperforms the state-of-the-art, 
showing performance speedup of 4.6\% in RFM-mitigated systems and reducing DRAM dynamic energy by 11.8\% in PRAC-mitigated systems.  ORAP uses 22.7\% less state in its largest form, and overcomes the limitations of previous DRAM-aware prefetchers by the targeted prefetching of deep access streams and issuing requests out-of-order to maximize rowbuffer hit rate and bank-level-parallelism. ORAP leverages large page sizes, large cache state, and modern memory systems to improve overall interactions between hardware prefetchers, caches, and DRAM to achieve significant performance improvement. While these contributions are considerable, more work can be done to improve this system. Other prefetchers may fare better alongside ORAP than HSD or Berti, and considerable improvements can still be made to ORAP's overall prefetcher usefulness.

\section{Acknowledgements}
Portions of this research were conducted with the advanced computing resources provided by Texas A\&M High Performance Research Computing.

The authors acknowledge the support from the Purdue Center for a Secure Microelectronics Ecosystem [CSME\#210205].

\bibliography{refs}

@inproceedings {zen4map,
author = {Patrick Jattke and Max Wipfli and Flavien Solt and Michele Marazzi and Matej B{\"o}lcskei and Kaveh Razavi},
title = {{ZenHammer}: Rowhammer Attacks on {AMD} Zen-based Platforms},
booktitle = {33rd USENIX Security Symposium (USENIX Security 24)},
year = {2024},
isbn = {978-1-939133-44-1},
address = {Philadelphia, PA},
pages = {1615--1633},
url = {https://www.usenix.org/conference/usenixsecurity24/presentation/jattke},
publisher = {USENIX Association},
month = aug
}

@inproceedings{minpage,
author = {Kaseridis, Dimitris and Stuecheli, Jeffrey and John, Lizy Kurian},
title = {Minimalist open-page: a DRAM page-mode scheduling policy for the many-core era},
year = {2011},
isbn = {9781450310536},
publisher = {Association for Computing Machinery},
address = {New York, NY, USA},
url = {https://doi.org/10.1145/2155620.2155624},
doi = {10.1145/2155620.2155624},
booktitle = {Proceedings of the 44th Annual IEEE/ACM International Symposium on Microarchitecture},
pages = {24–35},
numpages = {12},
location = {Porto Alegre, Brazil},
series = {MICRO-44}
}

@inproceedings{rubix,
author = {Saxena, Anish and Mathur, Saurav and Qureshi, Moinuddin},
title = {Rubix: Reducing the Overhead of Secure Rowhammer Mitigations via Randomized Line-to-Row Mapping},
year = {2024},
isbn = {9798400703850},
publisher = {Association for Computing Machinery},
address = {New York, NY, USA},
url = {https://doi.org/10.1145/3620665.3640404},
doi = {10.1145/3620665.3640404},
booktitle = {Proceedings of the 29th ACM International Conference on Architectural Support for Programming Languages and Operating Systems, Volume 2},
pages = {1014–1028},
numpages = {15},
location = {La Jolla, CA, USA},
series = {ASPLOS '24}
}

@inproceedings{rowhammer,
author = {Kim, Yoongu and Daly, Ross and Kim, Jeremie and Fallin, Chris and Lee, Ji Hye and Lee, Donghyuk and Wilkerson, Chris and Lai, Konrad and Mutlu, Onur},
title = {Flipping bits in memory without accessing them: an experimental study of DRAM disturbance errors},
year = {2014},
isbn = {9781479943944},
publisher = {IEEE Press},
booktitle = {Proceeding of the 41st Annual International Symposium on Computer Architecuture},
pages = {361–372},
numpages = {12},
location = {Minneapolis, Minnesota, USA},
series = {ISCA '14}
}

@INPROCEEDINGS{rhrevisit,
  author={Kim, Jeremie S. and Patel, Minesh and Yağlıkçı, A. Giray and Hassan, Hasan and Azizi, Roknoddin and Orosa, Lois and Mutlu, Onur},
  booktitle={2020 ACM/IEEE 47th Annual International Symposium on Computer Architecture (ISCA)}, 
  title={Revisiting RowHammer: An Experimental Analysis of Modern DRAM Devices and Mitigation Techniques}, 
  year={2020},
  volume={},
  number={},
  pages={638-651},
  doi={10.1109/ISCA45697.2020.00059}}

@inproceedings{rowpress,
author = {Luo, Haocong and Olgun, Ataberk and Ya\u{g}l\i{}k\c{c}\i{}, Abdullah Giray and Tu\u{g}rul, Yahya Can and Rhyner, Steve and Cavlak, Meryem Banu and Lindegger, Jo\"{e}l and Sadrosadati, Mohammad and Mutlu, Onur},
title = {RowPress: Amplifying Read Disturbance in Modern DRAM Chips},
year = {2023},
isbn = {9798400700958},
publisher = {Association for Computing Machinery},
address = {New York, NY, USA},
url = {https://doi.org/10.1145/3579371.3589063},
doi = {10.1145/3579371.3589063},
booktitle = {Proceedings of the 50th Annual International Symposium on Computer Architecture},
articleno = {28},
numpages = {18},
location = {Orlando, FL, USA},
series = {ISCA '23}
}

@misc{prac2,
      title={Understanding the Security Benefits and Overheads of Emerging Industry Solutions to DRAM Read Disturbance}, 
      author={Oğuzhan Canpolat and A. Giray Yağlıkçı and Geraldo F. Oliveira and Ataberk Olgun and Oğuz Ergin and Onur Mutlu},
      year={2024},
      eprint={2406.19094},
      archivePrefix={arXiv},
      primaryClass={cs.CR},
      url={https://arxiv.org/abs/2406.19094}, 
}

@misc{prac1,
      title={JESD79-5c: DDR5 SDRAM Standard}, 
      author={JEDEC},
      year={2024},
}

@misc{rfm,
      title={JESD79-5: DDR5 SDRAM Standard}, 
      author={JEDEC},
      year={2020},
}

@INPROCEEDINGS{exynos,
  author={Grayson, Brian and Rupley, Jeff and Zuraski, Gerald Zuraski and Quinnell, Eric and Jiménez, Daniel A. and Nakra, Tarun and Kitchin, Paul and Hensley, Ryan and Brekelbaum, Edward and Sinha, Vikas and Ghiya, Ankit},
  booktitle={2020 ACM/IEEE 47th Annual International Symposium on Computer Architecture (ISCA)}, 
  title={Evolution of the Samsung Exynos CPU Microarchitecture}, 
  year={2020},
  volume={},
  number={},
  pages={40-51},
  doi={10.1109/ISCA45697.2020.00015}}

@inproceedings{daampm,
author = {Ishii, Yasuo and Inaba, Mary and Hiraki, Kei},
title = {Unified memory optimizing architecture: memory subsystem control with a unified predictor},
year = {2012},
isbn = {9781450313162},
publisher = {Association for Computing Machinery},
address = {New York, NY, USA},
url = {https://doi.org/10.1145/2304576.2304614},
doi = {10.1145/2304576.2304614},
booktitle = {Proceedings of the 26th ACM International Conference on Supercomputing},
pages = {267–278},
numpages = {12},
location = {San Servolo Island, Venice, Italy},
series = {ICS '12}
}

@inproceedings{flipping-bits-in-memory,
author = {Kim, Yoongu and Daly, Ross and Kim, Jeremie and Fallin, Chris and Lee, Ji Hye and Lee, Donghyuk and Wilkerson, Chris and Lai, Konrad and Mutlu, Onur},
title = {Flipping bits in memory without accessing them: an experimental study of DRAM disturbance errors},
year = {2014},
isbn = {9781479943944},
publisher = {IEEE Press},
booktitle = {Proceeding of the 41st Annual International Symposium on Computer Architecuture},
pages = {361–372},
numpages = {12},
location = {Minneapolis, Minnesota, USA},
series = {ISCA '14}
}

@inproceedings{revisiting-rowhammer,
author = {Kim, Jeremie S. and Patel, Minesh and Ya\u{g}l\i{}k\c{c}\i{}, A. Giray and Hassan, Hasan and Azizi, Roknoddin and Orosa, Lois and Mutlu, Onur},
title = {Revisiting RowHammer: an experimental analysis of modern DRAM devices and mitigation techniques},
year = {2020},
isbn = {9781728146614},
publisher = {IEEE Press},
url = {https://doi.org/10.1109/ISCA45697.2020.00059},
doi = {10.1109/ISCA45697.2020.00059},
booktitle = {Proceedings of the ACM/IEEE 47th Annual International Symposium on Computer Architecture},
pages = {638–651},
numpages = {14},
location = {Virtual Event},
series = {ISCA '20}
}

@article{rowhammer-a-retrospective,
author = {Mutlu, Onur and Kim, Jeremie S.},
title = {RowHammer: A Retrospective},
year = {2020},
issue_date = {Aug. 2020},
publisher = {IEEE Press},
volume = {39},
number = {8},
issn = {0278-0070},
url = {https://doi.org/10.1109/TCAD.2019.2915318},
doi = {10.1109/TCAD.2019.2915318},
journal = {Trans. Comp.-Aided Des. Integ. Cir. Sys.},
month = aug,
pages = {1555–1571},
numpages = {17}
}

@INPROCEEDINGS{the-rowhammer-problem,
  author={Mutlu, Onur},
  booktitle={Design, Automation and Test in Europe Conference and Exhibition (DATE), 2017}, 
  title={The RowHammer problem and other issues we may face as memory becomes denser}, 
  year={2017},
  volume={},
  number={},
  pages={1116-1121},
  doi={10.23919/DATE.2017.7927156}}

@INPROCEEDINGS{rambleed,
  author={Kwong, Andrew and Genkin, Daniel and Gruss, Daniel and Yarom, Yuval},
  booktitle={2020 IEEE Symposium on Security and Privacy (SP)}, 
  title={RAMBleed: Reading Bits in Memory Without Accessing Them}, 
  year={2020},
  volume={},
  number={},
  pages={695-711},
  doi={10.1109/SP40000.2020.00020}}

@ARTICLE{memway,
  author={Xu, Lai and Yu, Rongwei and Wang, Lina and Liu, Weijie},
  journal={Tsinghua Science and Technology}, 
  title={Memway: in-memorywaylaying acceleration for practical rowhammer attacks against binaries}, 
  year={2019},
  volume={24},
  number={5},
  pages={535-545},
  doi={10.26599/TST.2018.9010134}}

@inproceedings{sledgehammer,
author = {Kang, Ingab and Wang, Walter and Kim, Jason and van Schaik, Stephan and Tobah, Youssef and Genkin, Daniel and Kwong, Andrew and Yarom, Yuval},
title = {SledgeHammer: amplifying rowhammer via bank-level parallelism},
year = {2024},
isbn = {978-1-939133-44-1},
publisher = {USENIX Association},
address = {USA},
booktitle = {Proceedings of the 33rd USENIX Conference on Security Symposium},
articleno = {90},
numpages = {18},
location = {Philadelphia, PA, USA},
series = {SEC '24}
}

@INPROCEEDINGS {mithril,
author = { Kim, Michael Jaemin and Park, Jaehyun and Park, Yeonhong and Doh, Wanju and Kim, Namhoon and Ham, Tae Jun and Lee, Jae W. and Ahn, Jung Ho },
booktitle = { 2022 IEEE International Symposium on High-Performance Computer Architecture (HPCA) },
title = {{ Mithril: Cooperative Row Hammer Protection on Commodity DRAM Leveraging Managed Refresh }},
year = {2022},
volume = {},
ISSN = {},
pages = {1156-1169},
doi = {10.1109/HPCA53966.2022.00088},
url = {https://doi.ieeecomputersociety.org/10.1109/HPCA53966.2022.00088},
publisher = {IEEE Computer Society},
address = {Los Alamitos, CA, USA},
month =apr}

@inproceedings{hydra,
author = {Qureshi, Moinuddin and Rohan, Aditya and Saileshwar, Gururaj and Nair, Prashant J.},
title = {Hydra: enabling low-overhead mitigation of row-hammer at ultra-low thresholds via hybrid tracking},
year = {2022},
isbn = {9781450386104},
publisher = {Association for Computing Machinery},
address = {New York, NY, USA},
url = {https://doi.org/10.1145/3470496.3527421},
doi = {10.1145/3470496.3527421},
booktitle = {Proceedings of the 49th Annual International Symposium on Computer Architecture},
pages = {699–710},
numpages = {12},
location = {New York, New York},
series = {ISCA '22}
}

@inproceedings{mint,
author = {Qureshi, Moinuddin and Qazi, Salman and Jaleel, Aamer},
title = {MINT: Securely Mitigating Rowhammer with a Minimalist in-DRAM Tracker},
year = {2024},
publisher = {IEEE Press},
url = {https://doi.org/10.1109/MICRO61859.2024.00071},
doi = {10.1109/MICRO61859.2024.00071},
booktitle = {Proceedings of the 2024 57th IEEE/ACM International Symposium on Microarchitecture},
pages = {899–914},
numpages = {16},
location = {Austin, TX, USA},
series = {MICRO '24}
}

@INPROCEEDINGS{blockhammer,
  author={Yağlikçi, A. Giray and Patel, Minesh and Kim, Jeremie S. and Azizi, Roknoddin and Olgun, Ataberk and Orosa, Lois and Hassan, Hasan and Park, Jisung and Kanellopoulos, Konstantinos and Shahroodi, Taha and Ghose, Saugata and Mutlu, Onur},
  booktitle={2021 IEEE International Symposium on High-Performance Computer Architecture (HPCA)}, 
  title={BlockHammer: Preventing RowHammer at Low Cost by Blacklisting Rapidly-Accessed DRAM Rows}, 
  year={2021},
  volume={},
  number={},
  pages={345-358},
  doi={10.1109/HPCA51647.2021.00037}}

@inproceedings{randomized-rowswap,
author = {Saileshwar, Gururaj and Wang, Bolin and Qureshi, Moinuddin and Nair, Prashant J.},
title = {Randomized row-swap: mitigating Row Hammer by breaking spatial correlation between aggressor and victim rows},
year = {2022},
isbn = {9781450392051},
publisher = {Association for Computing Machinery},
address = {New York, NY, USA},
url = {https://doi.org/10.1145/3503222.3507716},
doi = {10.1145/3503222.3507716},
booktitle = {Proceedings of the 27th ACM International Conference on Architectural Support for Programming Languages and Operating Systems},
pages = {1056–1069},
numpages = {14},
location = {Lausanne, Switzerland},
series = {ASPLOS '22}
}

@INPROCEEDINGS{graphene,
  author={Park, Yeonhong and Kwon, Woosuk and Lee, Eojin and Ham, Tae Jun and Ho Ahn, Jung and Lee, Jae W.},
  booktitle={2020 53rd Annual IEEE/ACM International Symposium on Microarchitecture (MICRO)}, 
  title={Graphene: Strong yet Lightweight Row Hammer Protection}, 
  year={2020},
  volume={},
  number={},
  pages={1-13},
  doi={10.1109/MICRO50266.2020.00014}}

@inproceedings{trespass,
	title = {{TRRespass}: {Exploiting} the {Many} {Sides} of {Target} {Row} {Refresh}},
	url = {Paper=https://download.vusec.net/papers/trrespass_sp20.pdf Slides=https://download.vusec.net/slides/trrespass_sp20.pdf Web=https://www.vusec.net/projects/trrespass Code=https://github.com/vusec/trrespass Press=https://bit.ly/2UXWKJ4},
	booktitle = {S\&{P}},
	author = {Frigo, Pietro and Vannacci, Emanuele and Hassan, Hasan and van der Veen, Victor and Mutlu, Onur and Giuffrida, Cristiano and Bos, Herbert and Razavi, Kaveh},
	month = may,
	year = {2020},
	note = {Best Paper Award, Pwnie Award for Most Innovative Research, IEEE Micro Top Picks Honorable Mention, DCSR Paper Award},
}

@INPROCEEDINGS{ppf,
  author={Bhatia, Eshan and Chacon, Gino and Pugsley, Seth and Teran, Elvira and Gratz, Paul V. and Jiménez, Daniel A.},
  booktitle={2019 ACM/IEEE 46th Annual International Symposium on Computer Architecture (ISCA)}, 
  title={Perceptron-Based Prefetch Filtering}, 
  year={2019},
  volume={},
  number={},
  pages={1-13},
  doi={}}

@INPROCEEDINGS{berti,
  author={Navarro-Torres, Agustín and Panda, Biswabandan and Alastruey-Benedé, Jesús and Ibáñez, Pablo and Viñals-Yúfera, Víctor and Ros, Alberto},
  booktitle={2022 55th IEEE/ACM International Symposium on Microarchitecture (MICRO)}, 
  title={Berti: an Accurate Local-Delta Data Prefetcher}, 
  year={2022},
  volume={},
  number={},
  pages={975-991},
  doi={10.1109/MICRO56248.2022.00072}}

@INPROCEEDINGS{bingo,
  author={Bakhshalipour, Mohammad and Shakerinava, Mehran and Lotfi-Kamran, Pejman and Sarbazi-Azad, Hamid},
  booktitle={2019 IEEE International Symposium on High Performance Computer Architecture (HPCA)}, 
  title={Bingo Spatial Data Prefetcher}, 
  year={2019},
  volume={},
  number={},
  pages={399-411},
  doi={10.1109/HPCA.2019.00053}}

@inproceedings{ampm,
author = {Ishii, Yasuo and Inaba, Mary and Hiraki, Kei},
title = {Access map pattern matching for data cache prefetch},
year = {2009},
isbn = {9781605584980},
publisher = {Association for Computing Machinery},
address = {New York, NY, USA},
url = {https://doi.org/10.1145/1542275.1542349},
doi = {10.1145/1542275.1542349},
booktitle = {Proceedings of the 23rd International Conference on Supercomputing},
pages = {499–500},
numpages = {2},
location = {Yorktown Heights, NY, USA},
series = {ICS '09}
}

@INPROCEEDINGS{bop,
  author={Michaud, Pierre},
  booktitle={2016 IEEE International Symposium on High Performance Computer Architecture (HPCA)}, 
  title={Best-offset hardware prefetching}, 
  year={2016},
  volume={},
  number={},
  pages={469-480},
  doi={10.1109/HPCA.2016.7446087}}

@INPROCEEDINGS{ipcp,
  author={Pakalapati, Samuel and Panda, Biswabandan},
  booktitle={2020 ACM/IEEE 47th Annual International Symposium on Computer Architecture (ISCA)}, 
  title={Bouquet of Instruction Pointers: Instruction Pointer Classifier-based Spatial Hardware Prefetching}, 
  year={2020},
  volume={},
  number={},
  pages={118-131},
  doi={10.1109/ISCA45697.2020.00021}}

@INPROCEEDINGS{aqua,
  author={Saxena, Anish and Saileshwar, Gururaj and Nair, Prashant J. and Qureshi, Moinuddin},
  booktitle={2022 55th IEEE/ACM International Symposium on Microarchitecture (MICRO)}, 
  title={AQUA: Scalable Rowhammer Mitigation by Quarantining Aggressor Rows at Runtime}, 
  year={2022},
  volume={},
  number={},
  pages={108-123},
  doi={10.1109/MICRO56248.2022.00022}}

@article{archshield,
author = {Nair, Prashant J. and Kim, Dae-Hyun and Qureshi, Moinuddin K.},
title = {ArchShield: architectural framework for assisting DRAM scaling by tolerating high error rates},
year = {2013},
issue_date = {June 2013},
publisher = {Association for Computing Machinery},
address = {New York, NY, USA},
volume = {41},
number = {3},
issn = {0163-5964},
url = {https://doi.org/10.1145/2508148.2485929},
doi = {10.1145/2508148.2485929},
journal = {SIGARCH Comput. Archit. News},
month = jun,
pages = {72–83},
numpages = {12}
}

@INPROCEEDINGS{autorfm,
  author={Qureshi, Moinuddin},
  booktitle={2025 IEEE International Symposium on High Performance Computer Architecture (HPCA)}, 
  title={AutoRFM: Scaling Low-Cost in-DRAM Trackers to Ultra-Low Rowhammer Thresholds}, 
  year={2025},
  volume={},
  number={},
  pages={991-1004},
  doi={10.1109/HPCA61900.2025.00078}}

@inproceedings{moat,
author = {Qureshi, Moinuddin and Qazi, Salman},
title = {MOAT: Securely Mitigating Rowhammer with Per-Row Activation Counters},
year = {2025},
isbn = {9798400706981},
publisher = {Association for Computing Machinery},
address = {New York, NY, USA},
url = {https://doi.org/10.1145/3669940.3707278},
doi = {10.1145/3669940.3707278},
booktitle = {Proceedings of the 30th ACM International Conference on Architectural Support for Programming Languages and Operating Systems, Volume 1},
pages = {698–714},
numpages = {17},
location = {Rotterdam, Netherlands},
series = {ASPLOS '25}
}

@inproceedings{when-mitigations-backfire,
author = {Woo, Jeonghyun and Qu, Joyce and Saileshwar, Gururaj and Nair, Prashant Jayaprakash},
title = {When Mitigations Backfire: Timing Channel Attacks and Defense for PRAC-Based RowHammer Mitigations},
year = {2025},
isbn = {9798400712616},
publisher = {Association for Computing Machinery},
address = {New York, NY, USA},
url = {https://doi.org/10.1145/3695053.3731007},
doi = {10.1145/3695053.3731007},
booktitle = {Proceedings of the 52nd Annual International Symposium on Computer Architecture},
pages = {739–756},
numpages = {18},
location = {
},
series = {ISCA '25}
}

@INPROCEEDINGS{spp,
  author={Kim, Jinchun and Pugsley, Seth H. and Gratz, Paul V. and Reddy, A.L. Narasimha and Wilkerson, Chris and Chishti, Zeshan},
  booktitle={2016 49th Annual IEEE/ACM International Symposium on Microarchitecture (MICRO)}, 
  title={Path confidence based lookahead prefetching}, 
  year={2016},
  volume={},
  number={},
  pages={1-12},
  doi={10.1109/MICRO.2016.7783763}}

@INPROCEEDINGS{asd,
  author={Hur, Ibrahim and Lin, Calvin},
  booktitle={2006 39th Annual IEEE/ACM International Symposium on Microarchitecture (MICRO'06)}, 
  title={Memory Prefetching Using Adaptive Stream Detection}, 
  year={2006},
  volume={},
  number={},
  pages={397-408},
  doi={10.1109/MICRO.2006.32}}

@inproceedings{kpc,
author = {Kim, Jinchun and Teran, Elvira and Gratz, Paul V. and Jim\'{e}nez, Daniel A. and Pugsley, Seth H. and Wilkerson, Chris},
title = {Kill the Program Counter: Reconstructing Program Behavior in the Processor Cache Hierarchy},
year = {2017},
isbn = {9781450344654},
publisher = {Association for Computing Machinery},
address = {New York, NY, USA},
url = {https://doi.org/10.1145/3037697.3037701},
doi = {10.1145/3037697.3037701},
booktitle = {Proceedings of the Twenty-Second International Conference on Architectural Support for Programming Languages and Operating Systems},
pages = {737–749},
numpages = {13},
location = {Xi'an, China},
series = {ASPLOS '17}
}

@inproceedings{pacman,
author = {Wu, Carole-Jean and Jaleel, Aamer and Martonosi, Margaret and Steely, Simon C. and Emer, Joel},
title = {PACMan: prefetch-aware cache management for high performance caching},
year = {2011},
isbn = {9781450310536},
publisher = {Association for Computing Machinery},
address = {New York, NY, USA},
url = {https://doi.org/10.1145/2155620.2155672},
doi = {10.1145/2155620.2155672},
booktitle = {Proceedings of the 44th Annual IEEE/ACM International Symposium on Microarchitecture},
pages = {442–453},
numpages = {12},
location = {Porto Alegre, Brazil},
series = {MICRO-44}
}

@INPROCEEDINGS{ship,
  author={Wu, Carole-Jean and Jaleel, Aamer and Hasenplaugh, Will and Martonosi, Margaret and Steely, Simon C. and Emer, Joel},
  booktitle={2011 44th Annual IEEE/ACM International Symposium on Microarchitecture (MICRO)}, 
  title={SHiP: Signature-based Hit Predictor for high performance caching}, 
  year={2011},
  volume={},
  number={},
  pages={430-441},
  doi={}}

@INPROCEEDINGS{ipstride,
  author={Fu, J.W.C. and Patel, J.H. and Janssens, B.L.},
  booktitle={[1992] Proceedings the 25th Annual International Symposium on Microarchitecture MICRO 25}, 
  title={Stride Directed Prefetching In Scalar Processors}, 
  year={1992},
  volume={},
  number={},
  pages={102-110},
  doi={10.1109/MICRO.1992.697004}}

@misc{champsim,
  title={The Championship Simulator: Architectural Simulation for Education and Competition},
  author={Gober, Nathan and Chacon, Gino and Wang, Lei and Gratz, Paul V. and Jim{\'e}nez, Daniel A. and Teran, Elvira and Pugsley, Seth and Kim, Jinchun},
  year={2022},
  eprint={2210.14324},
  eprinttype={arxiv},
  eprintclass={cs.AR}
}

@misc{ramulator2,
      title={Ramulator 2.0: A Modern, Modular, and Extensible DRAM Simulator}, 
      author={Haocong Luo and Yahya Can Tuğrul and F. Nisa Bostancı and Ataberk Olgun and A. Giray Yağlıkçı and Onur Mutlu},
      year={2023},
      eprint={2308.11030},
      archivePrefix={arXiv},
      primaryClass={cs.AR},
      url={https://arxiv.org/abs/2308.11030}, 
}

@inproceedings{simpoint,
author = {Perelman, Erez and Hamerly, Greg and Van Biesbrouck, Michael and Sherwood, Timothy and Calder, Brad},
title = {Using SimPoint for accurate and efficient simulation},
year = {2003},
isbn = {1581136641},
publisher = {Association for Computing Machinery},
address = {New York, NY, USA},
url = {https://doi.org/10.1145/781027.781076},
doi = {10.1145/781027.781076},
booktitle = {Proceedings of the 2003 ACM SIGMETRICS International Conference on Measurement and Modeling of Computer Systems},
pages = {318–319},
numpages = {2},
location = {San Diego, CA, USA},
series = {SIGMETRICS '03}
}

@misc{spec2017,
    title={SPEC CPU2017},
    key={Standard Performance Evaluation Corporation.},
    year={2025},
    url={https://www.spec.org/cpu2017},
}

@misc{spec2006,
    title={SPEC CPU2006},
    key={Standard Performance Evaluation Corporation.},
    year={2025},
    url={https://www.spec.org/cpu2006},
}

@misc{gaps,
      title={The GAP Benchmark Suite}, 
      author={Scott Beamer and Krste Asanović and David Patterson},
      year={2017},
      eprint={1508.03619},
      archivePrefix={arXiv},
      primaryClass={cs.DC},
      url={https://arxiv.org/abs/1508.03619}, 
}

@inproceedings{xsbench,
author = {Tramm, John and Siegel, Andrew and Islam, Tanzima and Schulz, Martin},
year = {2014},
month = {09},
pages = {},
title = {XSBench - The development and verification of a performance abstraction for Monte Carlo reactor analysis}
}

@misc {thp,
title={Transparent Huge Pages},
key={LWN},
year={2011},
url={http://lwn.net/Articles/423584/}
}

@misc {arm_pages,
title={Virtual memory support, armv4 and armv5},
key={Arm Holdings},
year={2025},
url={https://developer.arm.com/documentation/ddi0406/cb/Appendixes/ARMv4-and-ARMv5-Differences/System-level-memory-model/Virtual-memory-support}
}

@misc{amd_pages,
title={Database Tuning on Linux OS: Reference Guide for AMD EPYC™ 7002 Series Processors},
key={Advanced Micro Devices Inc.},
year={2019},
url={https://www.amd.com/content/dam/amd/en/documents/epyc-technical-docs/tuning-guides/2019-amd-epyc-7002-tg-linux-databse-56783_1_0.pdf}
}

@misc{amd_npages,
title={AMD-V™ Nested Paging – White Paper},
key={Advanced Micro Devices Inc.},
year={2008},
url={https://www.cse.iitd.ac.in/~sbansal/csl862-virt/2010/readings/NPT-WP-1%201-final-TM.pdf}
}

@misc{intel_pages,
title={Intel® 64 and IA-32 Architectures Software Developer Manuals},
key={Intel Corp.},
year={2025},
url={https://www.intel.com/content/www/us/en/developer/articles/technical/intel-sdm.html}
}

@inproceedings {270443,
author = {Juan Navarro and Sitaram Iyer and Alan Cox},
title = {Practical, Transparent Operating System Support for Superpages},
booktitle = {5th Symposium on Operating Systems Design and Implementation (OSDI 02)},
year = {2002},
address = {Boston, MA},
url = {https://www.usenix.org/conference/osdi-02/practical-transparent-operating-system-support-superpages},
publisher = {USENIX Association},
month = dec
}

@article{Kim_2025,
   title={Per-Row Activation Counting on Real Hardware: Demystifying Performance Overheads},
   volume={24},
   ISSN={2473-2575},
   url={http://dx.doi.org/10.1109/LCA.2025.3587293},
   DOI={10.1109/lca.2025.3587293},
   number={2},
   journal={IEEE Computer Architecture Letters},
   publisher={Institute of Electrical and Electronics Engineers (IEEE)},
   author={Kim, Jumin and Baek, Seungmin and Wi, Minbok and Nam, Hwayong and Kim, Michael Jaemin and Lee, Sukhan and Sohn, Kyomin and Ahn, Jung Ho},
   year={2025},
   month=jul, pages={217–220} }

@INPROCEEDINGS{blp_llc,
  author={Lee, Chang Joo and Narasiman, Veynu and Mutlu, Onur and Patt, Yale N.},
  booktitle={2009 42nd Annual IEEE/ACM International Symposium on Microarchitecture (MICRO)}, 
  title={Improving memory Bank-Level Parallelism in the presence of prefetching}, 
  year={2009},
  volume={},
  number={},
  pages={327-336},
  doi={}}

@misc{practical,
      title={PRACtical: Subarray-Level Counter Update and Bank-Level Recovery Isolation for Efficient PRAC Rowhammer Mitigation}, 
      author={Ravan Nazaraliyev and Saber Ganjisaffar and Nurlan Nazaraliyev and Nael Abu-Ghazaleh},
      year={2025},
      eprint={2507.18581},
      archivePrefix={arXiv},
      primaryClass={cs.AR},
      url={https://arxiv.org/abs/2507.18581}, 
}

@inbook{Bruce_2023, place={Palo Alto, California}, title={Arm Neoverse V2 platform: Leadership Performance and Power Efficiency for NextGeneration Cloud Computing, ML and HPC Workloads}, booktitle={2023 IEEE Hot Chips 35 symposium (HCS): 27-29 Aug. 2023}, publisher={IEEE}, author={Bruce, Magnus}, year={2023}}

@INPROCEEDINGS{pageprefetch,
  author={Vavouliotis, Georgios and Chacon, Gino and Alvarez, Lluc and Gratz, Paul V. and Jiménez, Daniel A. and Casas, Marc},
  booktitle={2022 55th IEEE/ACM International Symposium on Microarchitecture (MICRO)}, 
  title={Page Size Aware Cache Prefetching}, 
  year={2022},
  volume={},
  number={},
  pages={956-974},
  doi={10.1109/MICRO56248.2022.00070}}

@inproceedings{trident,
author = {Ram, Venkat Sri Sai and Panwar, Ashish and Basu, Arkaprava},
title = {Trident: Harnessing Architectural Resources for All Page Sizes in x86 Processors},
year = {2021},
isbn = {9781450385572},
publisher = {Association for Computing Machinery},
address = {New York, NY, USA},
url = {https://doi.org/10.1145/3466752.3480062},
doi = {10.1145/3466752.3480062},
booktitle = {MICRO-54: 54th Annual IEEE/ACM International Symposium on Microarchitecture},
pages = {1106–1120},
numpages = {15},
location = {Virtual Event, Greece},
series = {MICRO '21}
}

@article{vmware,
author = {Guo, Fei and Kim, Seongbeom and Baskakov, Yury and Banerjee, Ishan},
title = {Proactively Breaking Large Pages to Improve Memory Overcommitment Performance in VMware ESXi},
year = {2015},
issue_date = {July 2015},
publisher = {Association for Computing Machinery},
address = {New York, NY, USA},
volume = {50},
number = {7},
issn = {0362-1340},
url = {https://doi.org/10.1145/2817817.2731187},
doi = {10.1145/2817817.2731187},
journal = {SIGPLAN Not.},
month = mar,
pages = {39–51},
numpages = {13}
}

@misc{kaslr, title={Kernel address space layout randomization},url={https://lwn.net/Articles/569635/}, key={LWN}, year={2013}}

@misc{zen5, title={Ryzen 7 7800X3D - AMD},url={https://en.wikichip.org/wiki/amd/ryzen_7/7800x3d}, key={WikiChip}}
\bibliographystyle{IEEEtranS}

\end{document}